\documentclass[twocolumn]{aastex6}
\usepackage{amsmath}
\usepackage{amssymb}
\usepackage{subfigure}
\usepackage{graphicx}
\usepackage{color}

\usepackage{txfonts}

\usepackage{natbib}
\usepackage{threeparttable}
\usepackage{bm}

\turnoffedit

\fullcollaborationName{The Friends of AASTeX Collaboration}

\begin{document}


\title{The Galilean Satellites Formed Slowly from Pebbles}


\author{Yuhito Shibaike\altaffilmark{1,2}, Chris W. Ormel\altaffilmark{3,4}, Shigeru Ida\altaffilmark{1}, Satoshi Okuzumi\altaffilmark{5}, and Takanori Sasaki\altaffilmark{6}}







\altaffiltext{1}{Earth-Life Science Institute, Tokyo Institute of Technology, Meguro-ku, Tokyo 152-8550, Japan}
\altaffiltext{2}{Physikalisches Institut \& NCCR PlanetS, Universitaet Bern, CH-3012 Bern, Switzerland}
\altaffiltext{3}{Anton Pannekoek Institute, University of Amsterdam, Science Park 904,1090GE Amsterdam, The Netherlands}
\altaffiltext{4}{Department of Astronomy, Tsinghua University, Haidian DS 100084, Beijing, China}
\altaffiltext{5}{Department of Earth and Planetary Sciences, Tokyo Institute of Technology, Meguro-ku, Tokyo, 152-8551, Japan}
\altaffiltext{6}{Department of Astronomy, Kyoto University, Kitashirakawa-Oiwake-cho, Sakyo-ku, Kyoto, 606-8502, Japan}

\begin{abstract}
It is generally accepted that the four major (Galilean) satellites formed out of the gas disk that accompanied Jupiter's formation. However, understanding the specifics of the formation process is challenging as both small particles (pebbles) as well as the satellites are subject to fast migration processes. Here, we hypothesize a new scenario for the origin of the Galilean system, based on the capture of several planetesimal seeds and subsequent slow accretion of pebbles. To halt migration, we invoke an inner disk truncation radius, and other parameters are tuned for the model to match physical, dynamical, compositional, and structural constraints. In our scenario it is natural that Ganymede's mass is determined by pebble isolation. Our slow-pebble-accretion scenario then reproduces the following characteristics: (1) the mass of all the Galilean satellites; (2) the orbits of Io, Europa, and Ganymede captured in mutual 2:1 mean motion resonances; (3) the ice mass fractions of all the Galilean satellites; (4) the unique ice-rock partially differentiated Callisto and the complete differentiation of the other satellites. Our scenario is unique to simultaneously reproduce these disparate properties.
\end{abstract}




\section{Introduction} \label{introduction}
The four large satellites around Jupiter were discovered by Galileo Galilei over 400 years ago. The physical, dynamical, compositional, and structural properties of the satellite system are well known. The inner three satellites, Io, Europa, and Ganymede, are captured in mutual 2:1 mean motion resonances. The satellites' masses are similar at about $10^{-4}~M_{\rm J}$ (Jupiter mass). The satellites' ice mass fractions increase with their distance from Jupiter: Io is dry, Europa consist of $6-9~{\rm wt\%}$ ice, while Ganymede and Callisto have ice mass fractions of about $50~{\rm wt\%}$ \citep{kus05}. Uniquely, Callisto features an undifferentiated internal structure \citep{sch04}.

Any model describing the formation of the Galilean satellites, must obey these basic observational constraints. However, previous models have only explained parts of these characteristics and the models are inconsistent with each other. Given that the satellites reside in the same plane, it is natural to assume that they formed in a gas disk surrounding Jupiter—the circum-Jovian disk (CJD)—analogous to the formation of planets in circum-stellar disks (CSDs) \citep{can02,mos03a}. Previous works showed that if enough satellitesimals (km-sized bodies) exist in the disk, satellites with the current Galilean satellites’ mass can form by two-body collisions \citep{can06,sas10}. A problem with this scenario, however, is the formation of satellitesimals. Unless the CJD features pressure reversals or the gas flows outward on the midplane, the progenitor dust grains will not have had the time to conglomerate, because of the strong radial drift \citep{shi17,dra18}. Alternatively, planetesimals expelled from the CSD can be captured by the CJD by virtue of the large density of the latter disk. A problem with this planetesimal-capture scenario is, however, that the growing Jupiter and the opened gas gap push the planetesimals out from Jupiter's feeding zone, rendering the capture rate very low \citep{hay77,fuj13}.

Another problem formation models face is that of strong radial migration of planetesimals and satellites by aerodynamic and tidal forces. Previous works argued that today's Galilean satellites are the final survivors of a history in which an earlier generation of satellites repeatedly formed, but were lost because they migrated into Jupiter \citep{can06,sas10,ogi12,cil18}. This model is therefore very inefficient in its use of (already scarce) solid material supplied to the CJD. An alternative idea, which we invoke in our model, is to stop the inward migration by virtue of a cavity of the gas disk around Jupiter opened by a strong magnetic field of the planet. The resonance chain of the inner three satellites' orbits is actually consistent with a scenario where the cavity halted the migration of Io and then Europa and Ganymede were captured into the resonances one by one \citep{sas10,ogi12}.

The variation in the ice fraction also constitutes a formidable modeling challenge \citep{hel15a,hel15b}. In particular, the ice mass fraction of Europa ($6-9\%$)—rather small compared to Ganymede and Callisto—is hard to explain by the previous satellitesimal-accretion scenario because the rocky and icy satellitesimals must be radially mixed beyond the snowline \citep{dwy13}. Recently, a scenario where Europa accretes small icy particles, which dehydrated interior to the snowline, has been suggested to explain the small ice fraction \citep{ron17}. Another idea is to reproduce the fraction by accreting small particles by invoking the inward movement of the snowline (the place in the CJD corresponding to $T=160~{\rm K}$) in the final growth phases of the satellites \citep{can09}. A formation scenario for the TRAPPIST-1 system which has various ice mass fractions of planets may be able to apply to the Galilean satellites, where the seeds of the planets form around the snowline and accrete pebbles during their migration toward the star \citep{orm17a}.

Finally, the dichotomy between the differentiated Ganymede and the partially differentiated Callisto requires tuned conditions. Given that Ganymede and Callisto have similar mass and compositions, it is natural to likewise expect a similar thermal history for these neighboring satellites \citep{bar08}. Thermal evolution after their formation by the release of the gravitational energy during the differentiation and/or the impacts during the Late Heavy Bombardment might be able to make the dichotomy but still tuned conditions were needed \citep{fri83,bar10}.

Here we show that we have succeeded to construct a new scenario, ``Slow-Pebble-Accretion scenario'', which reproduces the physical, dynamical, compositional, and structural properties of the satellite system simultaneously and consistently. Naturally, in order to match the above constraints, our model is characterized by a number of parameters. Therefore, we do not argue that the Galilean satellite system is an inevitable result. It is a possible result, and with other choice of parameters, different satellite systems are reproduced. Nevertheless, a unique and strong point of our scenario is that the small amount of ice in Europa and the dichotomy between the internal structures of Ganymede and Callisto can be reproduced simultaneously by a single scenario with plausible assumptions.

The structure of this paper is as follows. First, in Section \ref{methods}, we explain the methods used in this work. We then show the results of our calculations in Section \ref{results} and discuss the validity of our assumptions In Section \ref{assessment}. These two sections are the key parts of this paper. We also add some discussion in Section \ref{discussion} and conclude our work in Section \ref{conclusions}.

\section{Methods}
\label{methods}
In order to create a scenario for the origin of the Galilean satellites consistent with almost all of the available constraints, we rely on combining a variety of modeling tools. We first provide an executive summary of the overall model (Section \ref{methodssummary}) and then discuss its elements in more detail.

\subsection{Model Summary}
\label{methodssummary}
\begin{figure*}
\plotone{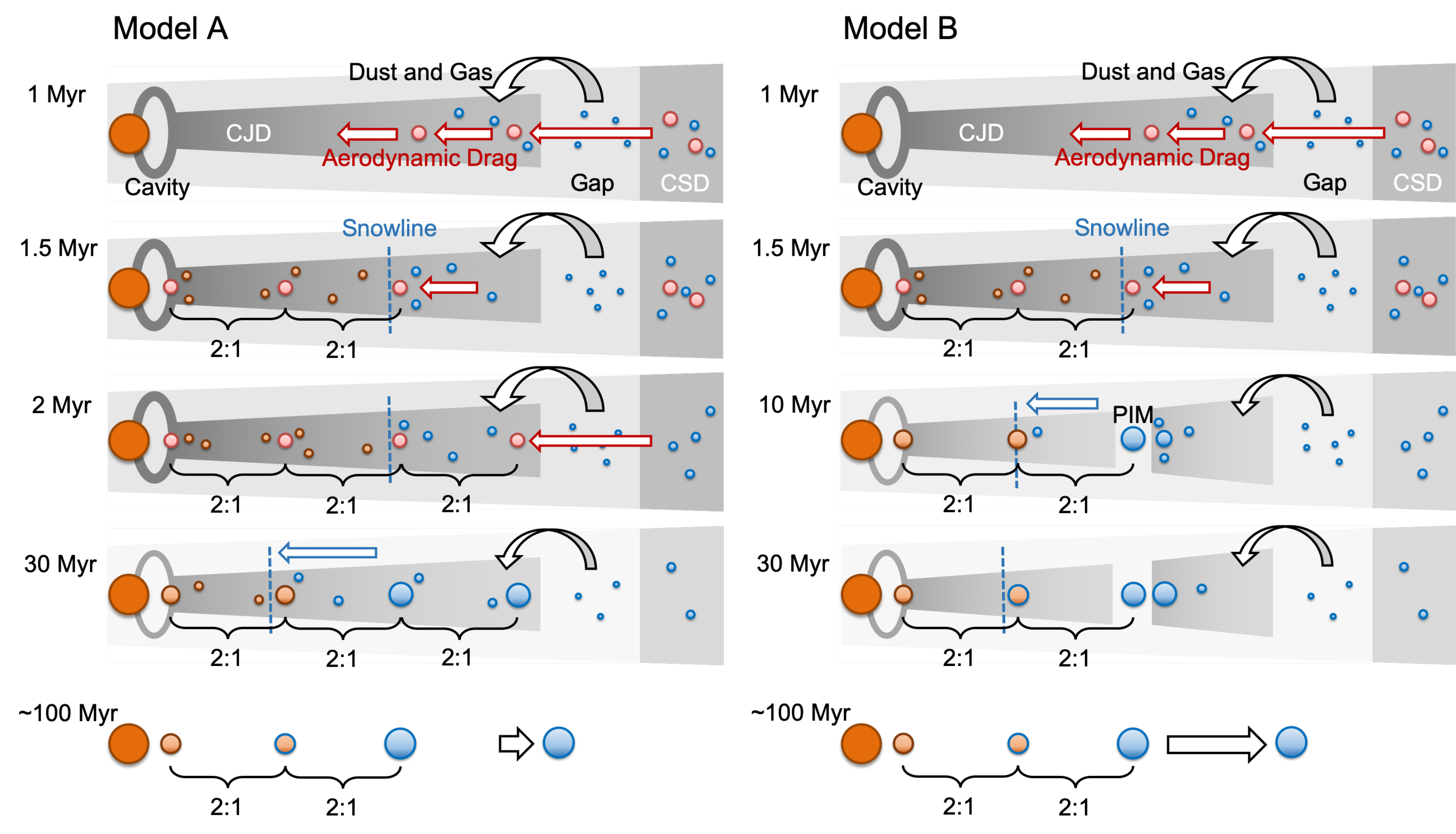}
\caption{{\bf Two models of the new formation scenario for the Galilean satellites.} {\bf Model A)} {\small $1~{\rm Myr}$) Jupiter grows to $\approx0.4~M_{\rm J}$. The gas accretion rate decreases due to a gap around the CJD and then an inner cavity around Jupiter opens. Pebbles drifting from the outer region of the CSD pile up at the pressure maximum of the gap. Only small dust particles coupled with gas are supplied to the CJD. Some planetesimals form from the pebbles at the pressure maximum. $1.5~{\rm Myr}$) Three planetesimals are captured by the CJD and migrate toward Jupiter by aerodynamic drag. The innermost one stops at the edge of the inner cavity and the two other planetesimals are captured into 2:1 mean motion resonances one by one. The position of the snow line is just inside the third satellite. $2~{\rm Myr}$) The fourth planetesimal is captured by the disk. It migrates inward quickly and is captured into a 2:1 resonance with the third one. The difference in capture time between the third and fourth planetesimals creates the dichotomy of their interior ice-rock differentiation. $30~{\rm Myr}$) The four planetesimals grow to the same sizes with the current Galilean satellites. The gas accretion rate decreases much because of the depletion of the parent CSD. The snowline then moves to just inside the second satellite (Europa) and small quantities of icy pebbles are accreted onto its outermost shell. $\sim100~{\rm Myr}$) The CJD has disappeared and the fourth satellite (Callisto) escapes from the resonance. The rock-metal differentiation occurs in Io by tidal heating and in Europa and Ganymede by long-lived radiogenic heating. {\bf Model B)} Until the third planetesimal is captured by the CJD, Model B is the same with Model A. $10~{\rm Myr}$) Ganymede reaches its PIM; a gas pressure maximum forms and drifting pebbles are trapped at the maximum, resulting in the termination of the growth of Io, Europa, and Ganymede. Europa accretes a small munber of icy pebbles before the termination of growth due to inward migration of the snowline. The seed of Callisto forms at the pressure maximum out of the trapped material. $30~{\rm Myr}$) Callisto grows to its current size, accreting the trapped material. $\sim100~{\rm Myr}$) The gas disk disappears and Callisto is scattered because its orbit is close to that of Ganymede.} \label{fig:idea}}
\end{figure*}

As we show later in Section \ref{PA}, the pebble isolation mass (PIM) in the CJD is close to the actual mass of Ganymede. Therefore, we investigate two models. In the first model, four planetesimals are captured, grow, and migrate (Model A). In the second model, three planetesimals are captured, the third satellite reaches pebble isolation, and the fourth satellite forms out of the pebbles trapped at the gas pressure maximum associated with reaching PIM (Model B). The methods used in the two models are the same except for the treatment of PIM and the way to calculate the growth of Callisto. In both the models, the captured planetesimals slowly accrete the particles drifting toward Jupiter, here referred to as pebbles ($\sim10~{\rm cm}$). The interiors of the satellites are mainly heated by the radiogenic decay of $^{26}$Al included in the accreted pebbles. Figure \ref{fig:idea} represents the two models of our slow-pebble-accretion scenario.

At first, we model the gas accretion rate as $\dot{M}_{\rm g}=0.2\exp(-(t-t_{\rm gap})/t_{\rm dep}))~M_{\rm J}~{\rm Myr}^{-1}$, where $t$, $t_{\rm gap}$, $t_{\rm dep}$, and $M_{\rm J}$ are the time after the formation of CAIs, gap opening time, gas depletion timescale of the CSD (we assume $t_{\rm dep}=3~{\rm Myr}$), and the current Jupiter mass $M_{\rm J}=1.90\times10^{30}~{\rm g}$, respectively. We consider a 1-D viscous accretion CJD. We assume that the gas mass flux is uniform in the CJD and it is equal to both the inflow mass flux to the disk and gas accretion rate to Jupiter, and the mass of Jupiter grows by $\dot{M}_{\rm g}$ from $0.4~M_{\rm J}$ at $t=t_{\rm gap}$ to $1.0~M_{\rm J}$ at the end of the calculation. We fix the position of the edge of the cavity $r_{\rm cav}$ at the current position of Io. On the distribution of the disk temperature in our model, the position of the snowline $r_{\rm snow}$ is at Ganymede's orbit when the gap opens and moves to Europa's orbit at the end of the disk evolution.

We calculate the evolution of dust particles in the CJD using a 1-D single-size analytical formula \citep{shi17}. They grow to pebbles and then drift to Jupiter because the gas disk rotates with sub-Kepler velocity which is slower than the rotating velocity of the pebbles so that they lose their angular momentum. We assume the dust inflow mass flux as $x\dot{M}_{\rm g}$, where $x$, the dust-to-gas accretion ratio, is a constant parameter with $x=0.0026$. The pebble mass flux $M_{\rm p}$ is equal to $x\dot{M}_{\rm g}$ at the outer edge of the disk, but reduces inwards because of filtering by satellites and evaporation of ice. There are only rocky pebbles inside the snowline and we assume that the pebble mass flux is halved. We also assume that the fragmentation of pebbles occur when their collisional velocity becomes faster than $5$ or $50~{\rm m~s^{-1}}$ for inside or outside the snowline, respectively \citep{wad09,wad13}. We calculate the disk midplane temperature as $T_{\rm d}=(3GM_{\rm cp}\dot{M}_{\rm g}/(8\pi\sigma_{\rm SB}r^{3}))^{1/4}g$, where $g=(3/8\tau+1/(4.8\tau))^{1/4}$, $\sigma_{\rm SB}$ and $G$ are the Stefan-Boltzmann constant and the Gravitational constant, respectively \citep{nak94}. The Rosseland mean opacity is $\tau=\kappa\Sigma_{\rm g}$, where the dust opacity is $\kappa=450r_{\rm gg}$ for $T_{\rm d}\geq160~{\rm K}$ and $\kappa=450(T_{\rm d}/160~{\rm K})^{2}r_{\rm gg}$ for $T_{\rm d}<160~{\rm K}$, where $r_{\rm gg}$ is the ratio of the surface density of grains affecting the disk temperature to gas.

We calculate the mass of the growing seeds as
\begin{equation}
M_{\rm s}(t)=\int_{t_{\rm cap}}^{t} \dot{M}_{\rm p}P_{\rm eff}~dt,
\label{Ms}
\end{equation}
where $t_{\rm cap}$ and $P_{\rm eff}$ are the capture time of the seeds and their pebble accretion efficiencies. The pebble accretion efficiency depends on the mass of the seed and the Stokes number of the pebbles around it \citep{orm18}. For Callisto in model B, which is located at a pressure bump, we need to modify Eq. (\ref{Ms}). See Section \ref{PA}.

We also calculate the migration of the seeds by aerodynamic drag and Type I migration which includes both inward and outward migration \citep{ada76,wei77,paa11b,ogi15}. We finally consider the capture into 2:1 or 3:2 mean motion resonances \citep{ogi13}. The orbit of Callisto is fixed in Model B at the gas pressure maximum created by Ganymede.

We calculate the surface temperature, $T_{\rm s}(R_{\rm s})$, with the radii of the seeds, $R_{\rm s}$, from the equilibrium of the emission and accretion heat of pebbles. We also calculate the cumulative heat of $^{26}$Al decay, $\Delta T_{\rm fin}(R)$, from when the pebbles (the ice mass fraction is 0.5) including $^{26}$Al have been accreted on the seeds' surface to the end of the formation, where $R$ is the distance from the center of the seeds. We can then estimate the final (i.e., maximum) internal temperature of the seeds at the point $R$ by $T_{\rm fin}(R)=T_{\rm s}(R)+\Delta T_{\rm fin}(R)$ \citep{bar08}. We do not include thermal diffusion, solid-state convection, and latent heat inside the seeds.

Below, we show the details of the methods.

\subsection{Circum-Jovian Disk model}\label{CJD}
After the gap formation, the gas accretion rate to Jupiter becomes much lower. In this work, we assume that the gas accretion rate is
\begin{equation}
\dot{M}_{\rm g}=\dot{M}_{\rm g,gap}\exp\left(-\dfrac{t-t_{\rm gap}}{t_{\rm dep}}\right) \;\;\; (t>t_{\rm gap}).
\label{Mdotg}
\end{equation}
We also assume that the initial gas accretion rate and the gas depletion timescale are $\dot{M}_{\rm g,gap}=0.2~M_{\rm J}~{\rm Myr}^{-1}$ and $t_{\rm dep}=3\times10^{6}~{\rm yr}$, where $M_{\rm J}$ is the current Jupiter mass. We set $t=0$ as the timing of the formation of CAIs. The mass of Jupiter grows by $\dot{M}_{\rm g}$ from $0.4~M_{\rm J}$ at $t=t_{\rm gap}$ to $1.0~M_{\rm J}$ at the end of the calculation. The accretion rate is equal to the inflow mass flux from the circum-stellar disk (CSD) to circum-Jovian disk (CJD) because we assume that the flow of gas (and pebbles) in whole region of the disk is semi-steady and all of the gas flow into the disk will be accreted by Jupiter eventually.

\begin{figure}
\epsscale{1.15}
\plotone{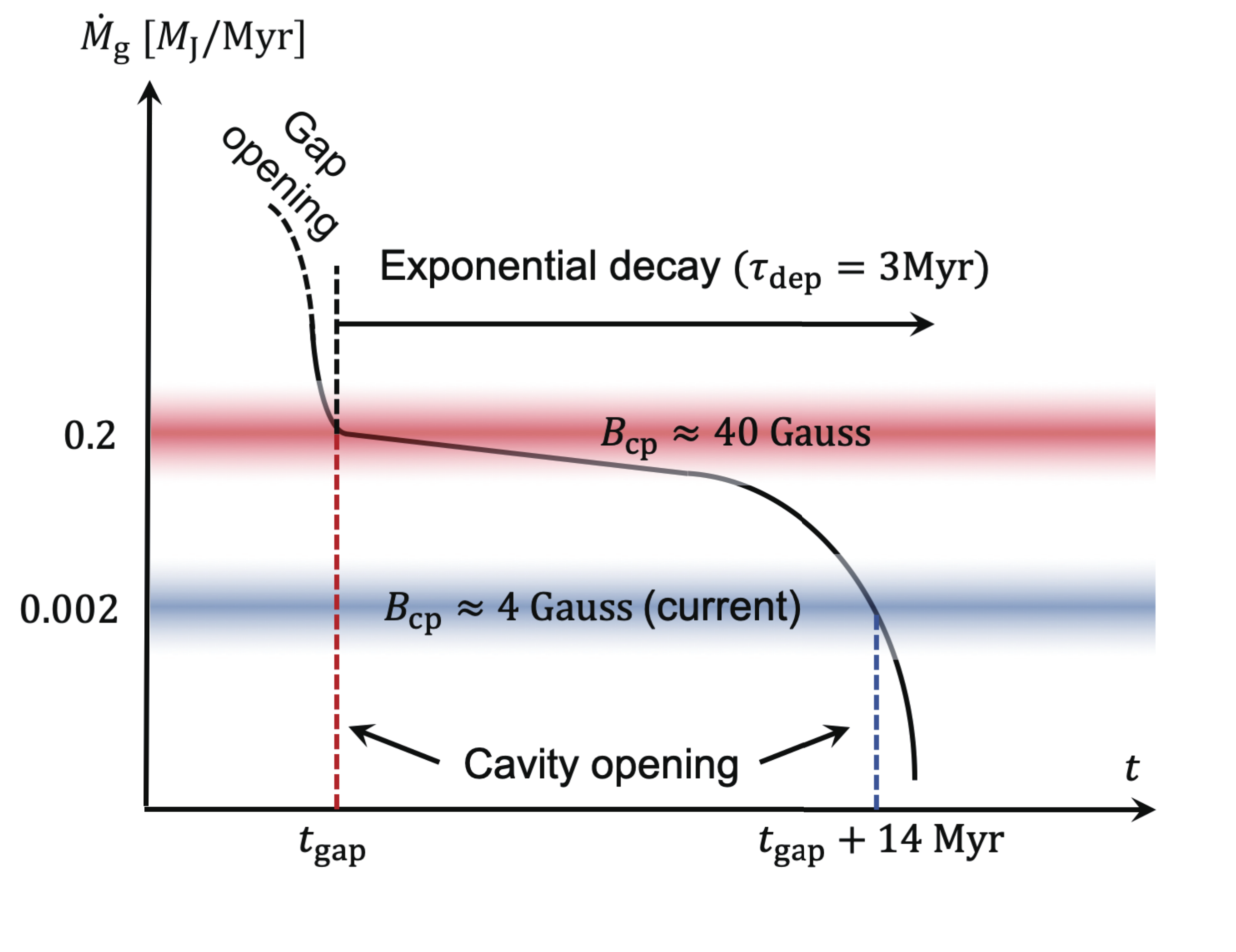}
\caption{{\bf Model of the gas accretion rate.} The black solid curve represents the evolution of the gas accretion rate in our model. After the gap opens at $t=t_{\rm gap}$, the accretion rate decreases exponentially. The red and blue regions represent the gas accretion rates required for the gap opening if the strength of Jupiter's magnetic field is $B_{\rm cp}\approx40$, or $4~{\rm Gauss}$ (i.e., the current strength), respectively. The red and blue dashed lines represent the time (after the formation of CAIs) that the inner cavity opens if $B_{\rm cp}\approx40$ or $4~{\rm Gauss}$, respectively. \label{fig:cases}}
\end{figure}

A cavity of the gas disk can open around Jupiter by the magnetic field of the planet. The timing that the inner cavity opens depends on the strength of the magnetic field. The position of the disk inner edge can be estimated from the balance of gas accretion and magnetic stress \citep{lov11,liu17}. If Jupiter has a dipole magnetic field, it is given by
\begin{equation}
\begin{split}
r_{\rm cav}=&\left(\dfrac{B_{\rm cp}^{4}R_{\rm cp}^{12}}{4GM_{\rm cp}\dot{M}_{\rm g}^{2}}\right)^{1/7} \\
=&1.07\left(\dfrac{B_{\rm cp}}{40~{\rm Gauss}}\right)^{4/7}\left(\dfrac{R_{\rm cp}}{R_{\rm J}}\right)^{12/7} \\
&\times\left(\dfrac{M_{\rm cp}}{0.4~M_{\rm J}}\right)^{-1/7}\left(\dfrac{\dot{M}_{\rm g}^{2}}{0.2~M_{\rm J}~{\rm Myr}^{-1}}\right)^{-2/7}~[R_{\rm J}], \\
\end{split}
\label{rcav}
\end{equation}
valid for $r_{\rm cav}<r_{\rm co}$ where $r_{\rm co}$ is the corotation radius (see Section \ref{cavity}). In the equation, $B_{\rm cp}$, $R_{\rm cp}$, $G$, and $M_{\rm cp}$ are the strength of the magnetic field of the central planet, radius of the central planet, gravitational constant, and mass of the central planet, respectively. Current Jupiter has a magnetic field and its strength on the surface of the equational region of the planet is $4.2~{\rm Gauss}$ \citep{con93}. Previous work, however, argued that the magnetic field was once stronger than the current one \citep{ste83,san04,chr09}. Therefore, we consider that the magnetic field is $\approx40~{\rm Gauss}$. In this case, the disk inner cavity and gap open at almost same time (substituting the Jupiter radius $R_{\rm J}=7.15\times10^{9}~{\rm cm}$ at $r_{\rm cav}$ for Eq. (\ref{rcav})). On the other hand, if the strength of the magnetic field is the same with the current one, the cavity only opens $14~{\rm Myr}$ later than the gap opening. In this case, any satellite would be consumed by proto-Jupiter ($t<14 {\rm Myr}$) and too little material would remain to form the Galilean satellites after gap opening. We summarize the evolution of $\dot{M}_{\rm g}$ in Figure \ref{fig:cases}. In our model, we assume that $r_{\rm cav}$ is fixed at the current Io's orbit, $5.89~R_{\rm J}$, for simplicity.

\begin{figure*}
\plotone{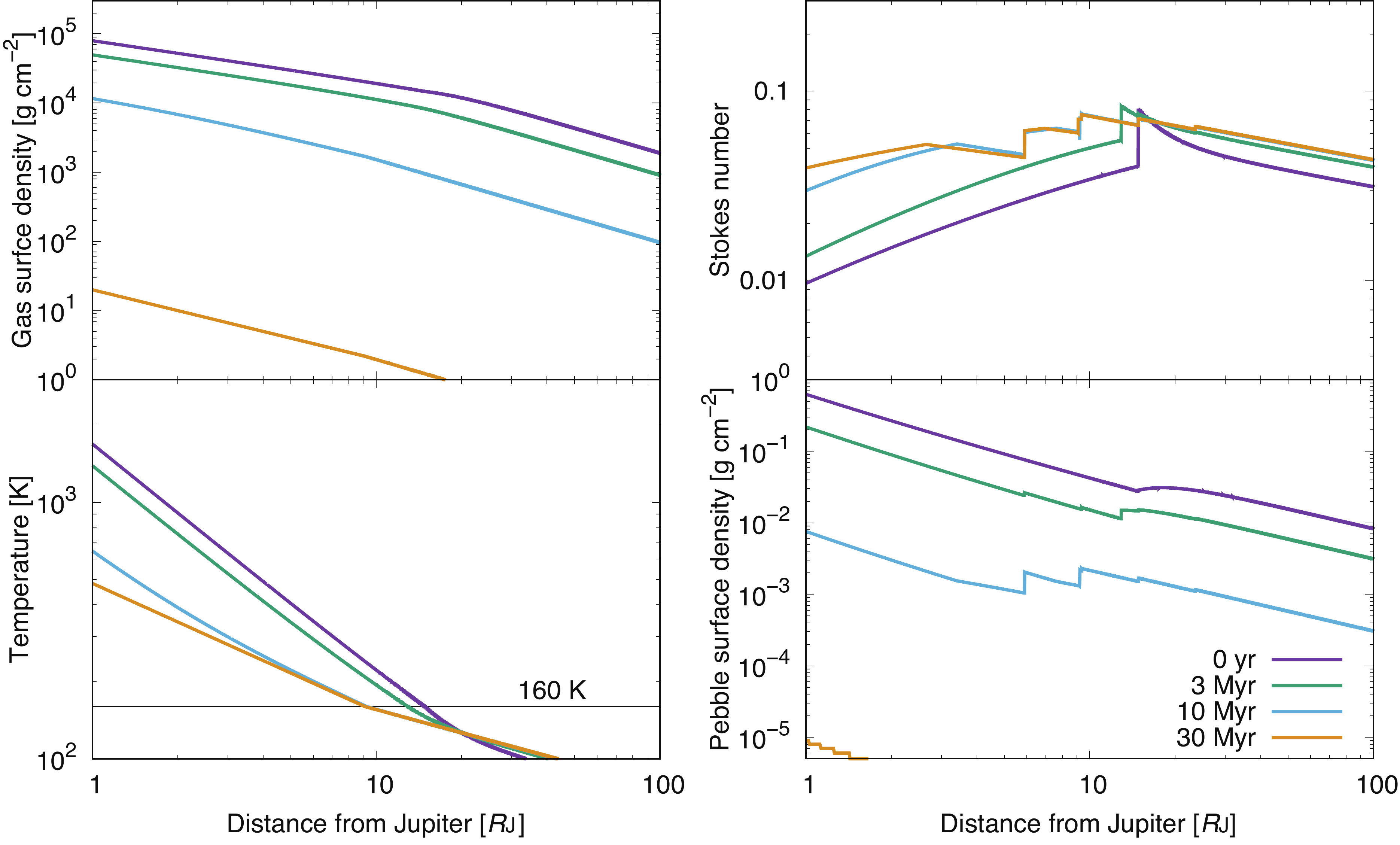}
\caption{{\bf Evolution of the circum-Jovian disk and the pebbles.} The left upper and lower panels represent the evolution of the gas surface density and the temperature of the midplane, respectively. The black horizontal line is the sublimation temperature of water ice, $160~{\rm K}$. The right upper and lower panels represent the evolution of the Stokes number of the drifting pebbles and the pebble surface density in Model A, respectively. The color variations of the both panels represent the time after the gap opens ($t-t_{\rm gap}$). \label{fig:disk}}
\end{figure*}

Based on the above gas accretion model, we calculate the evolution of the 1-D CJD. The gas surface density of the viscous accretion disk is,
\begin{equation}
\Sigma_{\rm g}=\dfrac{\dot{M}_{\rm g}\Omega_{\rm K}}{3\pi\alpha c_{\rm s}^{2}}.
\label{Sigmag}
\end{equation}
We assumed that the strength of turbulence is $\alpha=10^{-4}$ because  MRI is suppressed in the CJD \citep{fuj14}. The upper left panel of Figure \ref{fig:disk} represents the evolution of the gas surface density. The gas surface density becomes small as the gas accretion rate decreases.

We assume that the CJD is viscously heated. The gas temperature in the midplane of the viscous accretion disk is given by \citep{nak94},
\begin{equation}
T_{\rm d}=\left(\dfrac{3GM_{\rm cp}\dot{M}_{\rm g}}{8\pi\sigma_{\rm SB}r^{3}}\right)^{1/4}g,
\label{T4-b}
\end{equation}
where $\sigma_{\rm SB}$ is the Stefan-Boltzmann constant and,
\begin{equation}
g=\left(\dfrac{3}{8}\tau+\dfrac{1}{4.8\tau}\right)^{1/4}
\label{g-b}
\end{equation}
is a function of the Rosseland mean optical depth $\tau=\kappa\Sigma_{\rm g}$. In principle, the opacity $\kappa$ depends on the size distribution of solid particles. However, the size distribution cannot be predicted from the simple dust evolution model as employed in this study. Therefore, we just assume $r_{\rm gg}$, the ratio of the surface density of grains that affect the temperature to the gas surface density. Then, the opacity can be assumed as
\begin{equation}
\kappa=\begin{cases}
450r_{\rm gg} & T_{\rm d}\geq160~{\rm K} \\
450\left(T_{\rm d}/160~{\rm K}\right)^{2}r_{\rm gg} & T_{\rm d}<160~{\rm K}.
\label{kappa}
\end{cases}
\end{equation}

The lower left panel of Figure \ref{fig:disk} represents the evolution of the temperature when $r_{\rm gg}=1.7\times10^{-7}$. We choose this value to get the thermal condition that the snowline is just inside current Ganymede's orbit. The gradients of curves change at the snowline where the temperature is $160~{\rm K}$. The Rosseland mean optical depth $\tau$ becomes the smallest ($\sim1$) slightly outside the snowline. The temperature decreases as the gas accretion rate reduces. The temperature depends on this ratio $r_{\rm gg}$ and we discuss the dependence of the results on this parameter in detail in Section \ref{assessment}.

\subsection{Pebble Growth and Radial Drift} \label{pebblegrowth}
We calculate the distributions of the Stokes number and surface density of the drifting pebbles in the CJD. We only consider semi-steady conditions of pebbles because the evolution timescale of the pebbles is much shorter than those of the disk and satellites. In this case, the pebble mass flux in the CJD $\dot{M}_{\rm p}$ does not depend on the distance from Jupiter $r$. It is also equal to the mass flux of dust particles supplied from the parent CSD to the CJD,
\begin{equation}
\dot{M}_{\rm p}=x\dot{M}_{\rm g},
\label{Mdotp}
\end{equation}
where $x$ is the ratio of the dust-to-gas accretion rates. We treat this ratio as a parameter and assume that the ratio does not depend on time, for simplicity. However, the mass flux of pebbles drifting inside the snowline is smaller than that of outside because pebbles lose their $\rm{H_{2}O}$ ice inside the snowline. We assume that the rock mass fraction of pebbles outside the snowline is $m_{\rm r}=0.5$ so that $\dot{M}_{\rm p}$ inside the snowline becomes half of that of outside. We also consider the filtering effect by outer satellites.

Under these assumptions, we first calculate the Stokes number of the pebbles. When the Stokes number is determined by radial drift, it can be calculated by the following equation (modified version of Eq. (15) of \citet{shi17}),
\begin{equation}
\begin{split}
{\rm St_{p}}=0.23&\left(\dfrac{2}{3+2p+q}\right)^{4/5}\left(\dfrac{10}{18-39q}\right)^{2/5} \\
&\times\left(\dfrac{\dot{M}_{\rm p}/\dot{M}_{\rm g}}{0.003}\right)^{2/5}\left(\dfrac{\alpha}{10^{-4}}\right)^{1/5} \\
&\times\left(\dfrac{T_{\rm d}}{160~{\rm K}}\right)^{-2/5}\left(\dfrac{M_{\rm cp}}{1~M_{\rm J}}\right)^{2/5}\left(\dfrac{r}{10~R_{\rm J}}\right)^{-2/5}, \\
\end{split}
\label{stdrift}
\end{equation}
where $p$ and $q$ are the $r$ exponents of the gas surface density and temperature (i.e., $\Sigma_{\rm g}\propto r^{-p}$ and $T_{\rm d}\propto r^{-q}$).

However, fragmentation occurs if the collision velocity, in other words, the pebble-to-pebble relative velocity is too fast. This relative velocity is 
\begin{equation}
v_{\rm pp}=\sqrt{(v_{\rm r}/2)^{2}+v_{\rm t}^{2}},
\label{vpp}
\end{equation}
where $v_{\rm r}$ and $v_{\rm t}$ are the radial drift velocity of the pebbles and the relative velocity induced by turbulence, respectively. These two velocities are \citep{ada76,wei77,orm07},
\begin{equation}
v_{\rm r}=-2\dfrac{{\rm St_{p}}}{{\rm St_{p}}^{2}+1}\eta v_{\rm K},
\label{vrdash}
\end{equation}
where $v_{\rm K}=r\Omega_{\rm K}$ is the Kepler velocity, and,
\begin{equation}
v_{\rm t}=\sqrt{3\alpha}c_{\rm s}{\rm St_{p}}^{1/2}.
\label{vtdash}
\end{equation}
The ratio of the pressure gradient force to the gravity of Jupiter is,
\begin{equation}
\eta=-\dfrac{1}{2}\left(\dfrac{H_{\mathrm{g}}}{r}\right)^{2}\dfrac{\partial \ln{\rho_{\mathrm{g}}c_{\mathrm{s}}^{2}}}{\partial \ln{r}},
\label{eta}
\end{equation}
where $\rho_{\mathrm{g}}=\Sigma_{\mathrm{g}}/(\sqrt{2\pi}H_{\mathrm{g}})$ is the gas density at the midplane. If the Stokes number is limited by their fragmentation, it is
\begin{equation}
{\rm St_{p}}=\dfrac{-3\alpha c_{s}^{2}+\sqrt{9\alpha^{2}c_{s}^{4}+4\eta^{2}r^{2}\Omega_{\rm K}^{2}v_{\rm cr}^{2}}}{2\eta^{2}r^{2}\Omega_{\rm K}^{2}},
\label{stfrag}
\end{equation}
where $v_{\rm cr}$ is the critical fragmentation speed \citep{oku16}. This equation can be derived by substituting Eq. (\ref{vrdash}), (\ref{vtdash}), and $v_{\rm pp}=v_{\rm cr}$ for Eq. (\ref{vpp}).

Finally, the pebble surface density follows from the continuity equation,
\begin{equation}
\Sigma_{\rm p}=\dfrac{\dot{M}_{\rm p}}{2\pi rv_{\rm r}}.
\label{Sigmap}
\end{equation}

The right panels of Figure \ref{fig:disk} represent the evolution of the Stokes number and the surface density of the dust particles (i.e., pebbles). The stair around $10~R_{\rm J}$ is consistent with the position of the snowline. The snowline migrates inward because the temperature becomes lower as the gas accretion rate decreases. We also find that the Stokes number inside the snowline is smaller than that outside. Outside the snowline, the Stokes number is determined by drift (Eq. (\ref{stdrift})). While, inside the snowline, the Stokes number is determined by fragmentation (Eq. (\ref{stfrag})) because rocky particles are more fragile than icy ones \citep{wad09,wad13}. We assume that the critical fragmentation speeds of rocky and icy pebbles are $v_{\rm cr}=5$ and $50~{\rm m~s^{-1}}$, respectively. In the right panels of Figure \ref{fig:disk} the minor stairs reflect the accretion of pebbles by the planets. As time goes on and the satellites grow larger, the pebble accretion efficiency increases, resulting in larger jumps (see Section \ref{PA}). Although we do not consider the inner cavity in this calculation, the rocky pebbles should flow onto the planet with the gas because they are small enough to couple to the magnetospheric accretion flow of gas. Their dynamical timescale should be about the free-fall timescale, and the stopping time of pebbles is much smaller than it because the upper panel of Figure \ref{fig:disk} shows that the Stokes number is about $0.02$ around the cavity ($5.89~R_{\rm J}$).

\subsection{Pebble Accretion Efficiency} \label{PA}
According to recent N-body simulations, the pebble accretion efficiency is well-fitted by \citep{liu18}
\begin{equation}
P_{\rm eff}=\left\{\left(0.32\sqrt{\dfrac{\mu_{s}\Delta v/v_{\rm K}}{{\rm St_{p}}\eta^{2}}}\right)^{-2}+\left(0.39\dfrac{\mu_{\rm s}}{\eta h_{\rm p}}\right)^{-2} \right\}^{-1/2},
\label{Peff}
\end{equation}
where $\mu_{\rm s}=M_{\rm s}/M_{\rm cp}$ and $h_{\rm{p}}=H_{\rm p}/r$ are the satellite-to-central planet mass ratio and the pebble aspect ratio, respectively. This equation combines two regimes of pebble accretion, 2-D (the first term) and 3-D (the second term) limits. If the pebble accretion radius is larger than the pebble scale height $H_{\rm p}$, the first term is dominant. The pebble scale height can be derived analytically from the balance of their vertical sedimentation and diffusion \citep{you07},
\begin{equation}
H_{\rm{p}}=H_{\rm{g}}\left(1+\dfrac{{\rm St_{p}}}{\alpha}\dfrac{1+2{\rm St_{p}}}{1+{\rm St_{p}}}\right)^{-1/2},
\label{Hp}
\end{equation}
where $H_{\rm g}=c_{\rm s}/\Omega_{\rm K}$ is the gas scale height. The expression in the 2-D limit depends on the approach velocity of the pebbles $\Delta v$, which is given by the Keplerian shear in the disk or else the disk head wind. The approach velocity is
\begin{equation}
\Delta v/v_{\rm K}=0.52(\mu_{\rm s}{\rm St_{p}})^{1/3} + \eta\left\{1 + 5.7\left(\dfrac{\mu_{\rm s}}{\eta^{3}/{\rm St_{p}}}\right)\right\}^{-1},
\label{Deltav}
\end{equation}
where the first and second terns represent the shear and the head wind limits, respectively \citep{orm18}.

The pebble mass flux within an orbit of the seed is smaller than without because a fraction of drifted pebbles are captured by the embryo,
\begin{equation}
\dot{M}_{\rm p,in}=(1-P_{\rm eff})\dot{M}_{\rm p,out},
\label{Mdotpin}
\end{equation}
where $\dot{M}_{\rm p,in}$ and $\dot{M}_{\rm p,out}$ are the fluxes inside and outside the embryo, respectively. The growth timescale of the seeds is $t_{\rm grow}=M_{\rm s}/(dM_{\rm s}/dt)$.

Efficient pebble accretion occurs when the satellite seed mass exceeds a critical mass \citep{orm17b},
\begin{equation}
M_{\ast}=\dfrac{v_{\rm hw}^{3}t_{\rm stop}}{8G}=\dfrac{1}{8}\eta^{3}{\rm St_{p}}M_{\rm cp}
\label{Mstar}
\end{equation}
(settling regime). In this work, we assume that all seeds always accrete pebbles effectively. In our CJD model, we found that the gas aspect ratio $h_{\rm g}$ is about 0.1 in the whole disk regions and so $\eta\sim h_{\rm g}^{2}\sim10^{-2}$. The Stokes number of the pebbles is ${\rm St_{p}}\sim0.1$ (See Figure \ref{fig:disk}). The critical mass is then $M_{\ast}\sim10^{-8}M_{\rm cp}$, which is about 10 times smaller than the initial mass $M_{\rm ini}=3\times10^{23}{\rm g}$.

Pebble accretion stops if a gap structure forms around the seed \citep{lam14,ata18,bit18,joh19}. The critical mass, known as the pebble isolation mass (PIM), is $M_{\rm iso}\sim h_{\rm g}^{3}M_{\rm cp}$, which is close to the largest Galilean satellite -- Ganymede -- in our CJD model. Because of this similarity, we consider two scenarios: (i) Ganymede did not reach the PIM (Model A); and (ii) Ganymede's reached pebble isolation (Model B). In the latter case, we therefore define the PIM of Ganymede as its actual current mass, $M_{\rm G}=1.48\times10^{26}~{\rm g}$ (see also Section \ref{PIM}). We also assume that if the mass of a satellite reaches the PIM, the growth of the satellite stops and the pebble mass flux inside the satellite ($\dot{M}_{\rm p,in}$ in Eq.(\ref{Mdotpin})) becomes zero immediately.

At the gas pressure maximum, another satellite may form by pebble accretion because drifting pebbles pile up there. The growth rate of the satellite at the gas pressure maximum is given by,
\begin{equation}
\dfrac{dM_{\rm s}}{dt}=\min(R_{\rm col}\Sigma_{\rm p}\Omega_{\rm K} r^{2}, \dot{M}_{\rm p}),
\label{dMsdt}
\end{equation}
where $R_{\rm col}=2\pi \eta P_{\rm eff}$ is the dimensionless accretion rate. Note that $R_{\rm col}$ does not contain an explicit $\eta$-dependence. The approach velocity of pebbles in  $P_{\rm eff}$, $\Delta v/v_{\rm K}$, is then calculated by Eq. (\ref{Deltav}) without the headwind term (the second term of the right-hand side). \citet{kan18} argued that, in cases of planets in PPDs, when a planet reaches its PIM, the dust-to-gas surface density ratio outside the gap formed by the planet becomes about unity and the dust-rich region (ring) widens outward. Therefore, we assume $\Sigma_{\rm p}=\Sigma_{\rm g}$ here but we find that the $\dot{M}_{\rm p}$ term in Eq.(\ref{dMsdt}) nonetheless limits the growth of the satellite: as soon as pebbles arrive at the pressure bump, they are accreted.

\subsection{Satellite Internal Temperature} \label{sattemp}
We calculate the internal temperature of satellites to estimate the levels of differentiation. The surface temperature of satellites with radius $R_{\rm s}$, can be estimated by the following equation \citep{bar08}. The temperature $T_{\rm s}(R_{\rm s})$ is
\begin{equation}
\rho_{\rm s}C_{\rm p}(T_{\rm s}(R_{\rm s})-T_{\rm d})\dfrac{dR_{\rm s}}{dt}=\dfrac{1}{2}\dfrac{\dot{M}_{\rm s}u_{\rm i}^{2}}{4\pi R_{\rm s}^{2}}-\sigma_{\rm SB}(T_{\rm s}(R_{\rm s})^{4} - T_{\rm d}^{4}).
\label{Ts}
\end{equation}
This equation represents the balance of the energy in the thin layer of the pebbles accreted during the unit time on the surface of the satellites. The left-hand side is the energy necessary for heating the thin layer. The terms of the right-hand side are the collisional energy of the accreted pebbles and the emission form the surface.
The pebble-satellite collision velocity $u_{\rm i}$ can be estimated by
\begin{equation}
u_{\rm i}=\min({\sqrt{v_{\rm esc}^{2}+\Delta v^{2}}, v_{\rm set}}),
\label{ui}
\end{equation}
where $v_{\rm esc}=\sqrt{2GM_{\rm s}/R_{\rm s}}$ and $v_{\rm set}=g_{\rm s}t_{\rm stop}$ are the escape and settling velocities, respectively. The gravitational acceleration at the surface of the satellite is $g_{\rm s}=GM_{\rm s}/R_{\rm s}^{2}$.

Heating by $^{26}$Al is also very effective. The increase of the satellite internal temperature at $R$, distance from the center of the satellite, during the formation is
\begin{equation}
\begin{split}
\Delta T_{\rm fin}(R)=&\dfrac{1}{C_{\rm p}}\int_{t_{\rm acc}}^{t_{\rm fin}}m_{\rm r}q_{26}(t)dt \\
=&\dfrac{m_{\rm r}q_{26,0}}{C_{\rm p}\lambda_{26}}\exp(-\lambda_{26}t_{\rm acc}), \\
\end{split}
\label{DeltaTfin}
\end{equation}
where $t_{\rm acc}$ and $t_{\rm fin}$ are the time when the satellite radius $R_{\rm s}$ was equal to $R$ and the end of the formation, respectively \citep{bar08}. We assume the initial heating rate by $^{26}$Al, the specific heat, and the satellite density as $q_{26,0}=1.82\times10^{-7}~{\rm W~kg^{-1}}$, $C_{\rm p}=1400~{\rm J~kg^{-1}~K^{-1}}$, and $\rho_{\rm s}=1.9~{\rm g~cm^{-3}}$, which are the values in the table \citep{bar08}. Here, we assume that all $^{26}$Al heat has remained at the point until the end. Thermal diffusion can be ignored because it diffuses only $\sim10~{\rm km}$ in $10^{7}$ years. Solid-state convection can also be ignored because it can start at $t\sim10^{8}~{\rm year}$ \citep{bar08}. Latent heat is not included because the purpose of this estimation is to determine whether the satellites melt or not. The final (i.e., maximum) satellite internal temperature at $R$ can be then estimated by the sum of the two heating sources, the accretion heating and the $^{26}$Al heating,
\begin{equation}
T_{\rm fin}(R)= T_{\rm s}(R)+\Delta T_{\rm fin}(R).
\label{Tfin}
\end{equation}
Note that the $^{26}$Al heating is dominant (see Section \ref{Callisto}).

\subsection{Satellite Migration} \label{migration}
There are two main mechanisms that make satellites migrate in the disk, aerodynamic drag and Type I migration. We consider both mechanisms at the same time to calculate the migration of satellites. The aerodynamic drag migration velocity can be obtained by substituting the Stokes number of satellites, ${\rm St_{s}}$, for ${\rm St_{p}}$ in Eq. (\ref{vrdash}),
\begin{equation}
v_{\rm ad}=-2\dfrac{{\rm St_{s}}}{{\rm St_{s}}^{2}+1}\eta v_{\rm K}.
\label{vad}
\end{equation}
The Stokes number of satellites is,
\begin{equation}
{\rm St_{s}}=\dfrac{8}{3C_{\rm D}}\dfrac{\rho_{\rm s}R_{\rm s}}{\rho_{\rm g}\eta v_{\rm K}}\Omega_{\rm K}.
\label{Sts}
\end{equation}
The Type I migration velocity is,
\begin{equation}
v_{\rm t1}=b_{\rm t1}\left(\dfrac{M_{\rm s}}{M_{\rm cp}}\right)\left(\dfrac{\Sigma_{\rm g}r^{2}}{M_{\rm cp}}\right)\left(\dfrac{v_{\rm K}}{c_{\rm s}}\right)^{2}v_{\rm K},
\label{vt1}
\end{equation}
where $b_{\rm t1}$ is the migration constant, which depends on the distribution of the temperature and gas surface density of the CJD (see Eq. (10) of \citet{ogi15}). If $b_{\rm t1}$ is negative or positive, the satellite migrates inward or outward, respectively. The migration timescale is then $t_{\rm mig}=r/|v_{\rm ad}+v_{\rm mig}|$.

\subsection{Capture into Mean Motion Resonances} \label{resonance}
According to recent N-body simulation, the critical migration timescale for capture into mean motion resonances is \citep{ogi13},
\begin{equation}
t_{\rm crit}=C_{\rm MMR}\left(\dfrac{M_{\rm in}}{M_{\oplus}}\right)^{-4/3}\left(\dfrac{M_{\rm cp}}{M_{\odot}}\right)^{4/3}T_{\rm in},
\label{Tcrit}
\end{equation}
where $M_{\rm in}$, $M_{\oplus}$, and $M_{\odot}$ are the mass of the inner satellite, Earth, and Sun, respectively, and $T_{\rm in}$ is the orbital period of the inner satellite. If the migration timescale $t_{\rm mig}$ is longer than this critical timescale, the two bodies can be captured into the resonance. The capture coefficient $C_{\rm MMR}$ depends on the type of the resonance and the mass ratio of the two bodies. We summarize the values of $C_{\rm MMR}$ in Table \ref{tab:CMMR} cited from \citet{ogi13} which calculated numerical simulations \citep{ogi13}. Note that our 1-D model does not consider the eccentricity and inclination of the orbits.

\begin{table}
\centering
\caption{Capture coefficient $C_{\rm MMR}$ for 2:1 or 3:2 MMRs adopted from \citet{ogi13}}
\medskip
\begin{tabular}{lcc} \hline
Mass ratio & 2:1 & 3:2 \\ \hline \hline
$M_{\rm out}/M_{\rm in}\sim1$ & $1\times10^{6}$ & $2\times10^{5}$ \\
$M_{\rm out}/M_{\rm in}\lesssim0.1$ & $1\times10^{7}$ & $5\times10^{5}$ \\ \hline
\end{tabular}
\label{tab:CMMR}
\end{table}

\section{Results} \label{results}
We calculate the evolution of the mass and orbits of the satellites in the two models, Models A and B. We first show the results of Model A (Section \ref{modela}). We then also show the results of Model B and compare them with those of Model A (Section \ref{modelb}).

\subsection{Model A} \label{modela}
\begin{figure*}[h]
\plotone{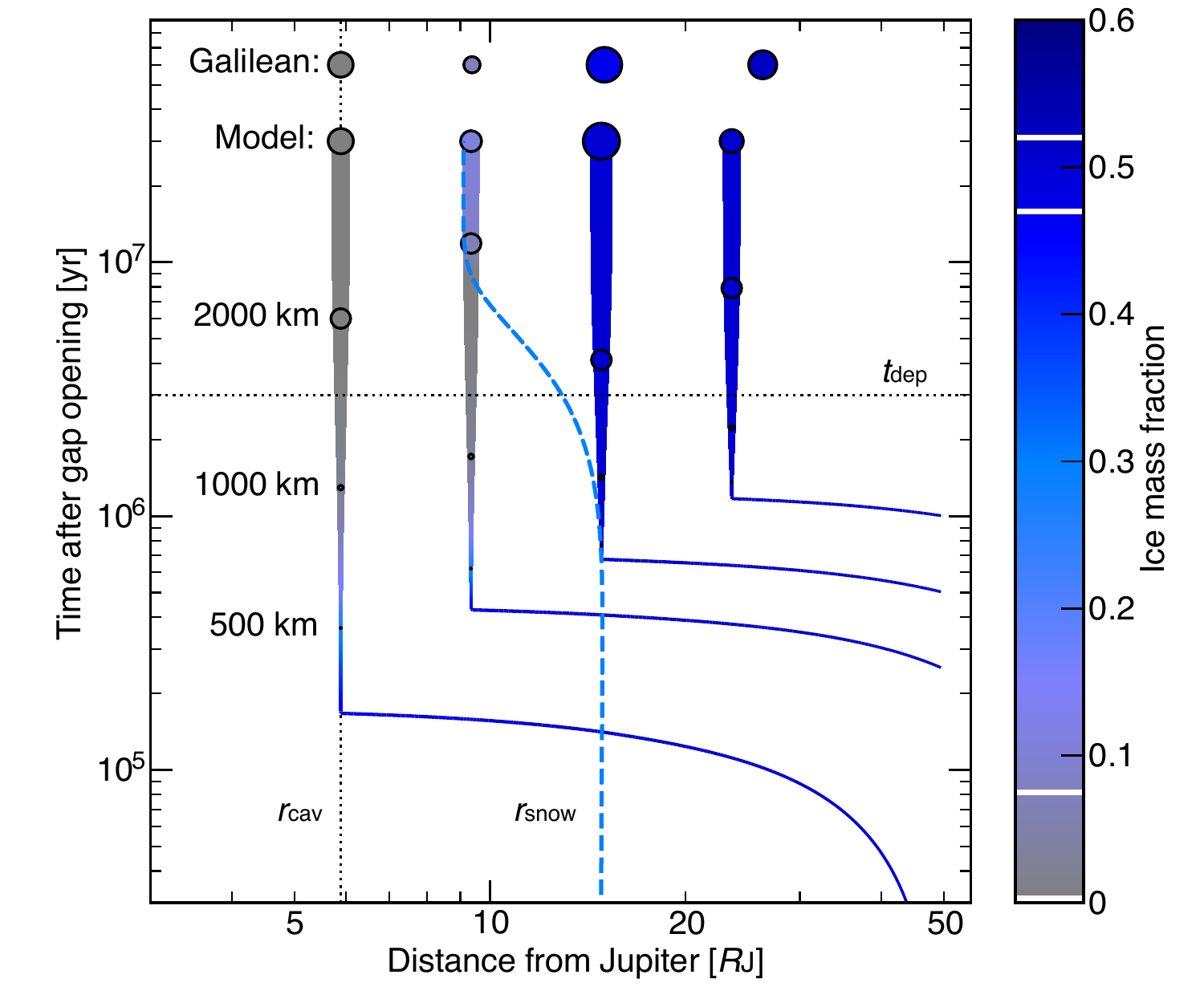}
\caption{{\bf Evolution of the four satellites (Model A).} The solid curves represent the positions of the evolving seeds (A1 to A4) at the time after the gap opens (i.e., $t-t_{\rm gap}$). The sizes of the circles represent the radii of the seeds and the current Galilean satellites. The color scales of the curves range from gray to dark blue for the increasing ice mass fractions of the seeds. The current ice mass fraction of the satellites (the mean values of the estimates by \citet{kus05}) are also shown as the color scale of the circles and the white lines in the column. The blue dashed curve represents the position of the snowline. The position of the edge of the inner cavity is fixed at the current orbit of Io (the vertical dotted line). The horizontal dotted line represents the gas depletion timescale of the CSD. The values of the final mass, positions, and ice mass fractions are shown in Table \ref{tab:seeds}. \label{fig:evol}}
\end{figure*}

We calculate the evolution of the mass and orbits of four planetesimals (Seeds A1 to A4) captured by the CJD one by one. We assume that the capture times of Seeds A1 to A4 are $t_{\rm cap}=1.0,~1.25,~1.5,~2.0~{\rm Myr}$, respectively. The disk condition is $x=0.0026$, $\alpha=10^{-4}$, and $r_{\rm gg}=1.7\times10^{-7}$. We also assume the initial mass and positions of the seeds as $M_{\rm s,start}=3\times10^{23}~{\rm g}$ and $r_{\rm s,start}=50~R_{\rm J}$, respectively. Figure \ref{fig:evol} represents the evolution of the size and orbits of the seeds, and Table \ref{tab:seeds} lists the final mass and positions.

Figure \ref{fig:evol} and Table \ref{tab:seeds} show that the mass (sizes) of the Galilean satellites can be reproduced very well. The dichotomy of the size between the inner and outer two satellites is created by the assumption that the pebble mass flux inside the snowline is half of that of outside because icy pebbles evaporate inside the snowline.

\begin{figure}[h]
\epsscale{1.15}
\plotone{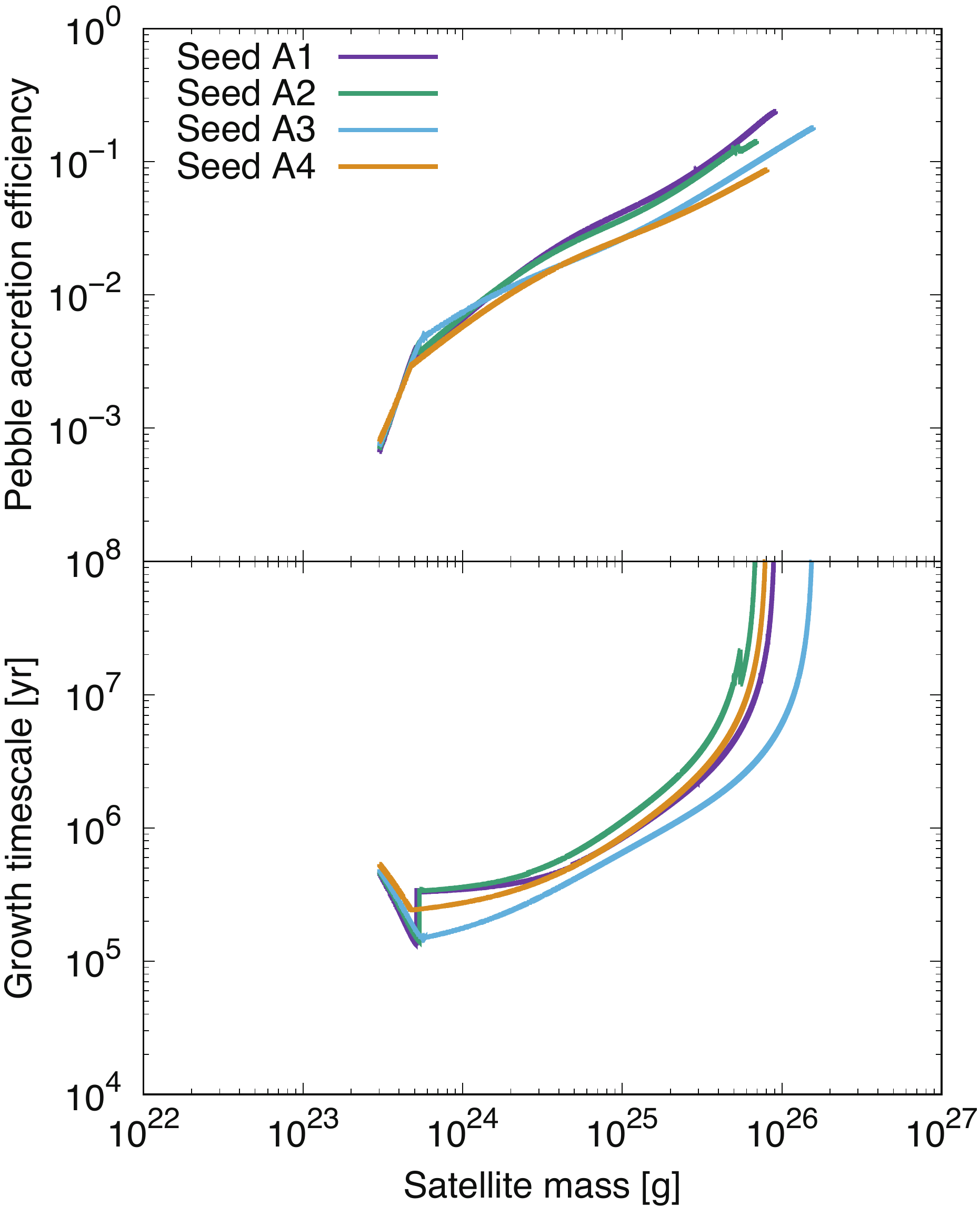}
\caption{{\bf Pebble accretion efficiency and growth timescale (Model A).} The purple, green, light blue, and orange curves in the upper panel represent the pebble accretion efficiencies of Seeds A1 to A4, respectively. The lower panel is the growth timescale of the seeds. \label{fig:peff}}
\end{figure}

The upper panel of Figure \ref{fig:peff} represents the pebble accretion efficiencies, $P_{\rm eff}$, of Seeds A1 to A4. All pebble accretion efficiencies are smaller than $\sim10\%$. The changes in the slope around $M_{\rm s}\approx5\times10^{23}~{\rm g}$ are caused by the stop of their migration due to the resonance traps. The mild changes in the gradients around $M_{\rm s}\approx2\times10^{24}~{\rm g}$ and $M_{\rm s}\approx10^{25}~{\rm g}$ correspond to the transition of the pebble accretion regimes from the 3-D to the 2-D and from the head wind to the shear regimes, respectively. The Stokes number inside of the snowline is smaller than that of outside by a factor of 2-3 (Figure \ref{fig:disk}), which makes the pebble accretion efficiencies of Io and Europa larger than that of Ganymede and Callisto in the 2-D regime because $P_{\rm eff}$ is a decreasing function (see Eqs. (\ref{Peff}), (\ref{Deltav})). The lower panel represents the growth timescale of the seeds, $t_{\rm grow}$. The timescale increases with mass and finally shoots up as the CJD disperses. It also shows that Ganymede grows fastest because the pebble mass flux becomes half after they passing the snowline, which has more impact than the reduction in ${\rm St}$ in the 3-D and the 2-D-shear regimes (see Eqs. (\ref{Ms}), (\ref{Peff}), (\ref{Hp}), and (\ref{Deltav})). The small jump around $M_{\rm s}\approx5\times10^{25}~{\rm g}$ on Europa's curve is caused by the crossing of the snowline. The filtering effects are mildly important for the growth timescale. Note that their mass does not reach the pebble isolation mass, $M_{\rm iso}$ (Eq. (\ref{Miso})).

Figure \ref{fig:evol} also shows that all seeds migrate quickly ($<3\times10^{5}~{\rm year}$) by aerodynamic drag (not by Type I migration) and are captured into 2:1 resonances one by one from the inner ones. After the seeds are captured into the resonances, they grow by pebble accretion without migration and keep their orbits on the current ones. The position of Seed A4, on the other hand, differs from the real orbit of Callisto. Seed A4 is also captured into a 2:1 resonance with Seed A3. However, Callisto's orbit may be able to expand rapidly escaping from the resonance by the resonant dynamical tide works \citep{ful16}.

Each Galilean satellite has the different ice mass fraction and, in particular, the low ice mass fraction of Europa is very unique (Table \ref{tab:seeds}). We find that this low ice mass fraction of Europa ($6-9\%$) can be reproduced by the migration of the snowline at the final phase of the formation (Table \ref{tab:seeds}). Figure \ref{fig:evol} shows that Seed 2 (Europa) accretes icy pebbles after $10~{\rm Myr}$. Although the ice mass fraction strongly depends on the disk temperature profile, there is a disk condition which is suitable for reproducing the ice mass fractions of all the Galilean satellites. In our model ice sublimation occurs instantaneously at the snowline, in contrast to \citet{ron17}, where pebbles only gradually lose their ice. Europa also naturally acquired an icy surface on top of a rocky interior, because the satellite accretes dry pebbles before accreting ice-rich pebbles.

We find that, in order to avoid differentiation of Callisto by $^{26}$Al heat, its seed must be captured by the disk late enough. The solid curves of Figure \ref{fig:temp} represent the internal temperature of Ganymede (Seed A3, light blue) and Callisto (Seed A4, orange). The first one is higher than the melting point of Callisto (black) and the second one is lower than it. This means that Callisto does not melt but Ganymede may melt by $^{26}$Al heat. The dichotomy of their internal ice-rock differentiation can be created by the difference in their capture time, $0.5~{\rm Myr}$, because the half-life of $^{26}$Al is $0.717~{\rm Myr}$. The long growth timescale ($\sim10^{7}~{\rm yr}$) is the reason why such different capture time is allowed. If the growth timescale is shorter and the difference in the capture time is the same, the final mass of Ganymede and Callisto would end up too large. Indeed, it is difficult to make the dichotomy between the internal structures of the two satellites by the classical Canup-Ward formation scenario where the growth timescale is $<10^{6}~{\rm yr}$ \citep{bar08}. Note that once rock-metal cores form in Europa and Ganymede, they can also differentiate metallic cores by long-lived radiogenic heating \citep{spo98}, and Io can be differentiated completely by tidal heating after its formation with its current orbit \citep{pea79}. See also Section \ref{RTI} discussing the subsequent evolution of Ganymede's internal structure.

\subsection{Model B} \label{modelb}
\begin{figure*}[h]
\plotone{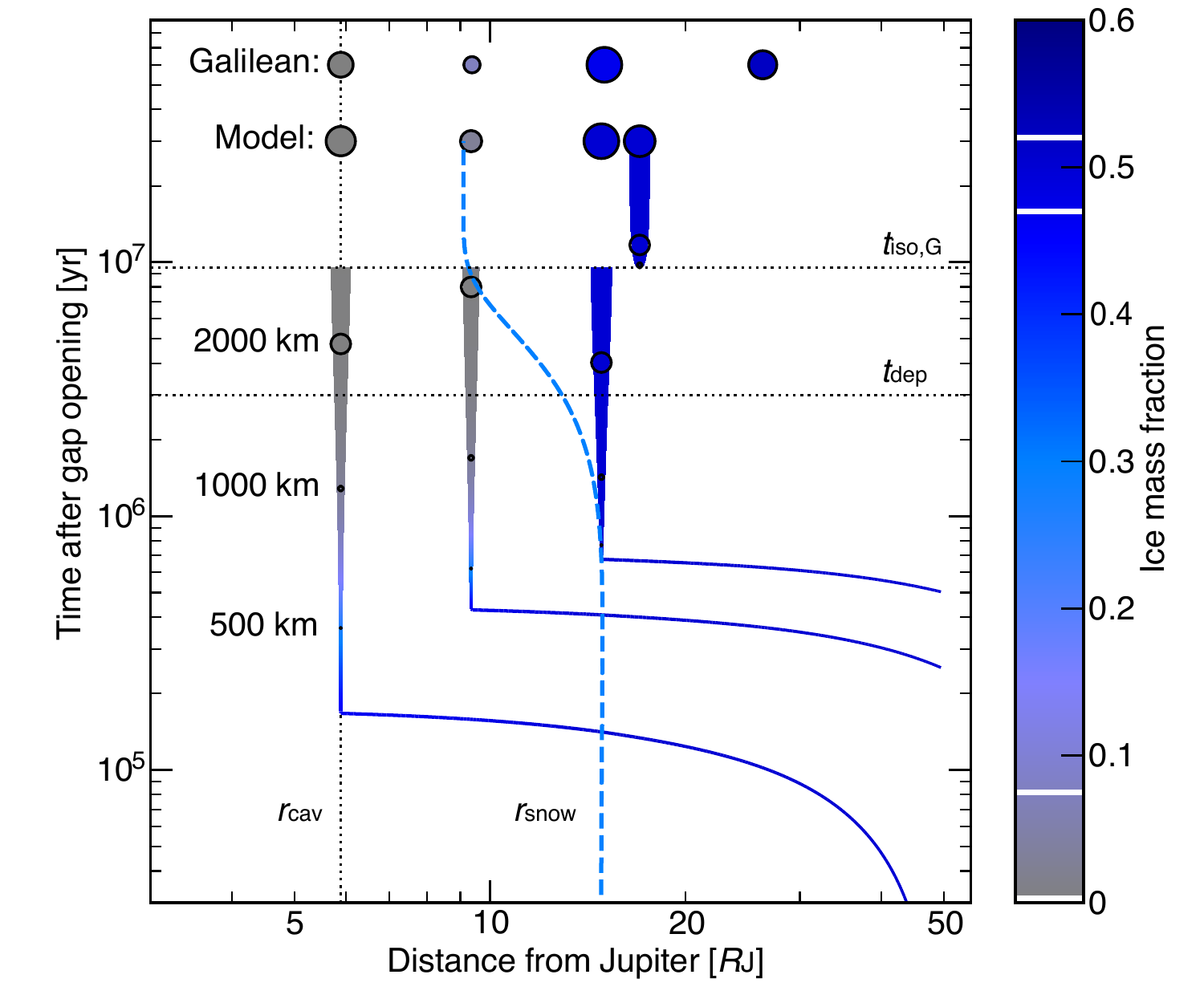}
\caption{{\bf Evolution of the four satellites (Model B).} Same as Figure \ref{fig:evol} but the evolution of Seeds B1 to B4 are shown. The time that Seed B3 reaches its PIM is $t_{\rm iso,G}=9.54~{\rm Myr}$ after the gap opening. The values of the final mass, positions, and ice mass fractions are shown in Table \ref{tab:seeds}.\label{fig:evolb}}
\end{figure*}

We then calculate the evolution of the satellites in a situation that three planetesimals (Seeds B1 to B3) are captured at first and the fourth seed (B4) forms and grows at the gas pressure bump of Seed B3. We assume that Seed B4 is put at its orbit and it starts to grow by pebble accretion when the mass of Seed B3 reaches its PIM, assumed to be equal to Ganymede's mass. We fix the orbit of Seed 4 at the gas pressure maximum formed by Seed B3, $r=17.0~R_{\rm J}$ (see Appendix \ref{maximum}). The other initial conditions are identical to those in Model A. Figure \ref{fig:evolb} represents the evolution of the size and orbits of the seeds, and Table \ref{tab:seeds} lists the final mass and positions.

Figure \ref{fig:evolb} and Table \ref{tab:seeds} show that the mass of all the satellites is reproduced by Model B as well. We find that the evolution of the inner three satellites is identical to Model A, except for the filtering effect of Callisto. In Model A, Seed A4 starts to decrease the pebble mass flux inside its orbit at $t-t_{\rm gap}=1~{\rm Myr}$. Conversely, Seed B4 appears only at $t-t_{\rm gap}=9.54~{\rm Myr}$ when Seed B3 reaches its PIM. Therefore, the inner satellites in Model B become slightly more massive than those in Model A. The growth of Callisto stops by the disappearance of the disk. Although we do not include the PIM of Callisto, the satellite should not reach pebble isolation under our assumption that the mass of the largest satellite, Ganymede, is defined as its PIM and PIM barely depends on the distance from Jupiter.

Figure \ref{fig:evolb} shows that the inner three satellites are captured into 2:1 MMRs also in Model B. On the other hand, in Model B, we assume that the orbit of Callisto is at the gas pressure maximum made by Ganymede, $r=17.0~R_{\rm J}$. The distance between the two orbits, $2.2~R_{\rm J}$, is consistent with $5.0~r_{\rm H,G}$, where $r_{\rm H,G}=(M_{\rm G}/(3M_{\rm J}))^{1/3}r_{\rm G}$ is the Hill radius of Ganymede, and $r_{\rm G}=14.8~R_{\rm J}$ is the orbital radius of Ganymede (in Model B). This distance is slightly wider than $2\sqrt{3}~r_{\rm H,G}$ but much narrower than $10~r_{\rm H,G}$. Therefore, Callisto is so close to Ganymede that the satellite should be scattered after the gas disk disappears and may reproduce its orbit which is not captured in any resonance \citep{cha96}.

We find that it is also possible in Model B to reproduce the small amount of ice on Europa (Table \ref{tab:seeds}). Figure \ref{fig:evolb} shows that Seed B2 accretes icy pebbles only at the end of the growth because Seed B3 reaches its PIM and traps drifting pebbles at its orbit just after Seed B2 starts to accrete icy pebbles. As a result, Europa's ice mass fraction is only $2.9\%$, smaller than that in Model A but not much smaller than the actual value, $6-9\%$ \citep{kus05}.

\begin{figure}[h]
\epsscale{1.15}
\plotone{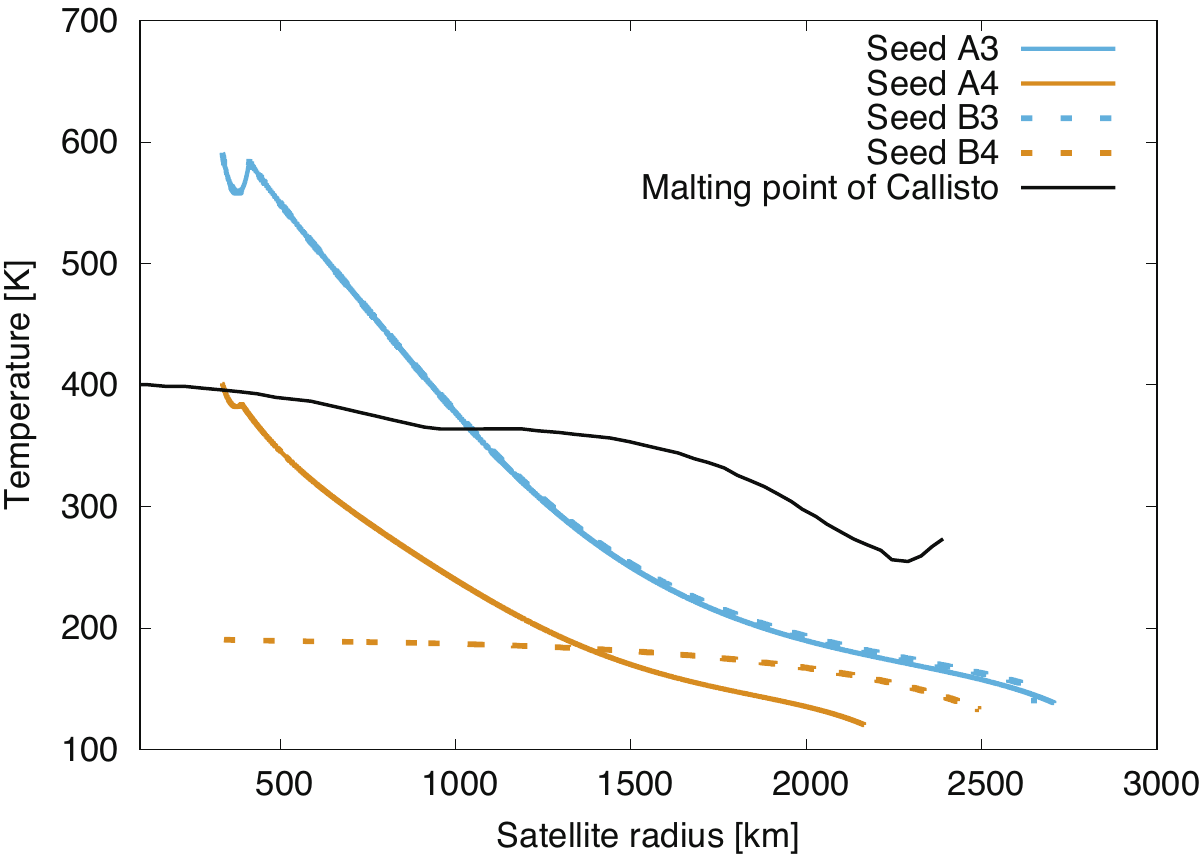}
\caption{{\bf Internal temperature of Ganymede and Callisto.} The light blue and orange curves represent the final ($t-t_{\rm gap}=30~{\rm Myr}$) internal temperature of Ganymede (Seeds A3 and B3) and Callisto (Seed A4 and B4), respectively. The solid and dashed curves are the results of Model A and B, respectively. The black curve is the melting point of Callisto cited by Figure 5 in \citet{bar08}. \label{fig:temp}}
\end{figure}

The dashed curves in Figure \ref{fig:temp} represent the internal temperature of Seeds B3 (light blue) and B4 (orange). The internal temperature of Seed B3 is similar to Seed A3 and higher than the melting point of Callisto (black). The internal temperature of Seed B4 is lower than the melting point in the whole interior region and is almost uniform. This is because $^{26}$Al radiogenic decay does not heat the satellite since Seed B4 starts to grow at later time ($9.54~{\rm Myr}$), which is naturally achieved in Model B. Moreover, the difference between the internal temperatures of Ganymede and Callisto is larger than that of Model A. We also find that the temperature is dominated by accretion heating and depends on the dust-to-gas accretion rate ratio.

\begin{table*}[h]
\centering
\caption{Final properties of the satellites}
\medskip
\begin{threeparttable}
\begin{tabular}{lccc} \hline
Seeds & Mass [$10^{25}$ g] & Orbital Position [$R_{\rm J}$] & Ice Mass Fraction [wt\%] \\ \hline \hline
Seed A1 & 9.14 & 5.89 & 0.28 \\
Seed A2 & 6.98 & 9.35 & 11 \\ 
Seed A3 & 15.8 &14.8 & 50 \\
Seed A4 & 8.08 & 23.6 & 50 \\ \hline
Seed B1 & 11.5 & 5.89 & 0.22 \\
Seed B2 & 7.12 & 9.35 & 2.9 \\ 
Seed B3 & 14.8\tnote{1} &14.8 & 50 \\
Seed B4 & 12.3 & 17.0 & 50 \\ \hline \hline
Io & 8.93 & 5.89 & 0\tnote{2} \\
Europa & 4.80 & 9.38 & 6-9\tnote{2} \\ 
Ganymede & 14.8 & 15.0 & 46-48\tnote{2} \\
Callisto & 10.8 & 26.3 & 49-55\tnote{2} \\ \hline
\end{tabular}
\begin{tablenotes}
\item[1] by definition
\item[2] Estimated current ice mass fractions by \citet{kus05}.
\end{tablenotes}
\end{threeparttable}
\label{tab:seeds}
\end{table*}

\section{Assessment} \label{assessment}
Although our new slow-pebble-accretion scenario reproduces most of the important properties of the current Galilean satellites -- mass, orbits, ice mass fractions, and internal structures -- many parameters needed to be tuned in order to meet the constraints. We summarize the key assumptions in Table \ref{tab:assumptions} and discuss each of them in the following paragraphs. The first column lists the key assumptions of our model and the second and third columns describe the motivations for them. Before discussing the sensitivity of the results to these parameters in detail, we would like to remark that the many constraints available for the Jovian system necessarily compels us to adopt specific parameter choices. Nonetheless, most of assumptions are supported by (or not inconsistent with) previous predictions or observations, and (as we will see below) our model would cope with a modest level of parameter variation. Our scenario, for the first time, reproduces most of the characteristics of the Galilean satellites simultaneously and consistently. Here, we show the detailed assessment of each of the assumptions.

\begin{table*}[h]
\centering
\caption{Summary of the key assumptions and their validity}
\medskip
\begin{threeparttable}
\begin{tabular}{lll} \hline
Key assumptions & Reproduced characteristics & Supporting predictions or observations \\ \hline \hline
Gap opening time & Melted/unmelted Ganymede \& Callisto & Early formation of Jupiter \\
$t_{\rm gap}=1.0~{\rm Myr}$ & & \\ \hline
Gas depletion timescale & - & Lifetime of PPDs \\
$t_{\rm dep}=3~{\rm Myr}$ & & \\ \hline
Width of the magnetospheric cavity & Position of Io & Stronger magnetic field of young Jupiter \\
$r_{\rm cav}=5.89~R_{\rm J}$ & & Larger radius of young Jupiter \\
 & & Photophoresis in the CJD \\ \hline
Strength of turbulence & Mass of all the satellites & Inactivity of MRI in CPDs \\ 
$\alpha=10^{-4}$ & Ice mass fractions of all the satellites & \\ \hline
Grain-to-gas surface density ratio & Ice mass fractions of all the satellites & Opacity of pebbles \\
$r_{\rm gg}=1.7\times10^{-7}$ & Mass of all the satellites & Dissipation of accretion energy at high altitude \\ \hline
Dust-to-gas accretion rate ratio & Mass of all the satellites & A small amount of dust-supply \\
$x=0.0026$ & & \\ \hline
Number of seeds & Four large Jovian satellites & Low likelihood of planetesimal-capture \\
Four & & \\ \hline
Eccentricity of orbits & Resonances of the inner three satellites & Long-lived prograde captured orbits \\
Zero & & \\ \hline
Initial mass of satellites (seeds) & Resonances of the inner three satellites & Long-lived prograde captured orbits \\
$M_{\rm s,start}=3\times10^{23}~{\rm g}$ & & Critical mass of pebble accretion's start \\ \hline
Capture time & Melted/unmelted Ganymede \& Callisto & - \\
$t_{\rm cap}=1.0, 1.25, 1.5, 2.0\tnote{1}~{\rm Myr}$ & & \\ \hline
Critical fragmentation speed & Mass of all the satellites & Previous numerical simulations \\
$v_{\rm cr}=5, 50~{\rm m~s^{-1}}$ & & Previous experiments \\ \hline
 PIM of Ganymede\tnote{2} & Mass of all the satellites & Previous numerical simulations \\
$M_{\rm iso,G}=M_{\rm G}=1.48\times10^{26}~{\rm g}$ & Melted/unmelted Ganymede & \\ \hline
\end{tabular}
\begin{tablenotes}
\item[1] Only in Model A
\item[2] Only in Model B
\end{tablenotes}
\end{threeparttable}
\label{tab:assumptions}
\end{table*}

\subsection{Early Formation of Jupiter} \label{Jupiter}
In our scenario, Jupiter has to grow so large that the gap structure opens and the pebbles can be trapped at the gas pressure maximum around Jupiter at $1~{\rm Myr}$ after the formation of the calcium-aluminum-rich inclusions (CAIs) to make the dichotomy of the internal structures of Ganymede and Callisto by $^{26}$Al heat. This is consistent with a resent ``early Jupiter formation'' scenario showing that solid materials in the solar system were spatially separated by that time \citep{kru17}.

\subsection{Gas Inflow to the CJD} \label{inflow}
We have assumed that the mass of young Jupiter is $0.4~M_{\rm J}$ when the gap forms and it becomes $1~M_{\rm J}$ at the end of the calculation. In our models, the mass flux of the gas inflow to the CJD, $\dot{M}_{\rm g}$, decreases as an exponential decay with the timescale of $t_{\rm dep}=3~{\rm Myr}$, which is consistent with the observations of the lifetime of the CSDs \citep{hai01}. It is generally believed that even after a gap forms, the gas accretion onto the planet (i.e., the gas inflow to the CPD) continues and its flux is determined by the gas surface density of the CPD inside the gap \citep{tan16}. Therefore, our assumption about the gas inflow is plausible. We note that, however, if the viscous gas accretion rate onto the Sun is lower than the photoevaporation rate, the gas surface density of the CSD and the gas accretion rate onto the CJD quickly decrease \citet{ale06a, ale06b}. In this case, the migration of snowline must be quick and the supply mechanism of ice to Europa may not work well, but the accretion of partially dehydrated pebbles argued in \citet{ron17} may be an alternative way.

\subsection{Inner Cavity} \label{cavity}
In our model, we have fixed the position of the disk inner edge as the current orbit of Io, $5.89~R_{\rm J}$, for simplicity, but this assumption is plausible. It is generally accepted that Jupiter's magnetic field was stronger than the current one, resulting in magnetospheric accretion and opening of the cavity in the latter phase of Jupiter's formation i.e., after the gap formed around the CJD \citep{ste83,san04,chr09}. The position of the edge of the cavity can be estimated from the balance of the gas accretion rate and magnetic stress by Jupiter's magnetic field (Eq. (\ref{rcav})). Since the gas accretion keeps decreasing after the gap opens, the edge moves outward and the innermost seed (Io) may also migrates together \citep{liu17}. Then the edge should stop at the corotation radius. It is generally believed that Jupiter was larger than today \citep{bur97,for11}. Conceivably, its spin frequency was lower and the corotation radius was located further than the current, $r_{\rm co}=2.25~R_{\rm J}$.

\begin{figure}
\epsscale{1.15}
\plotone{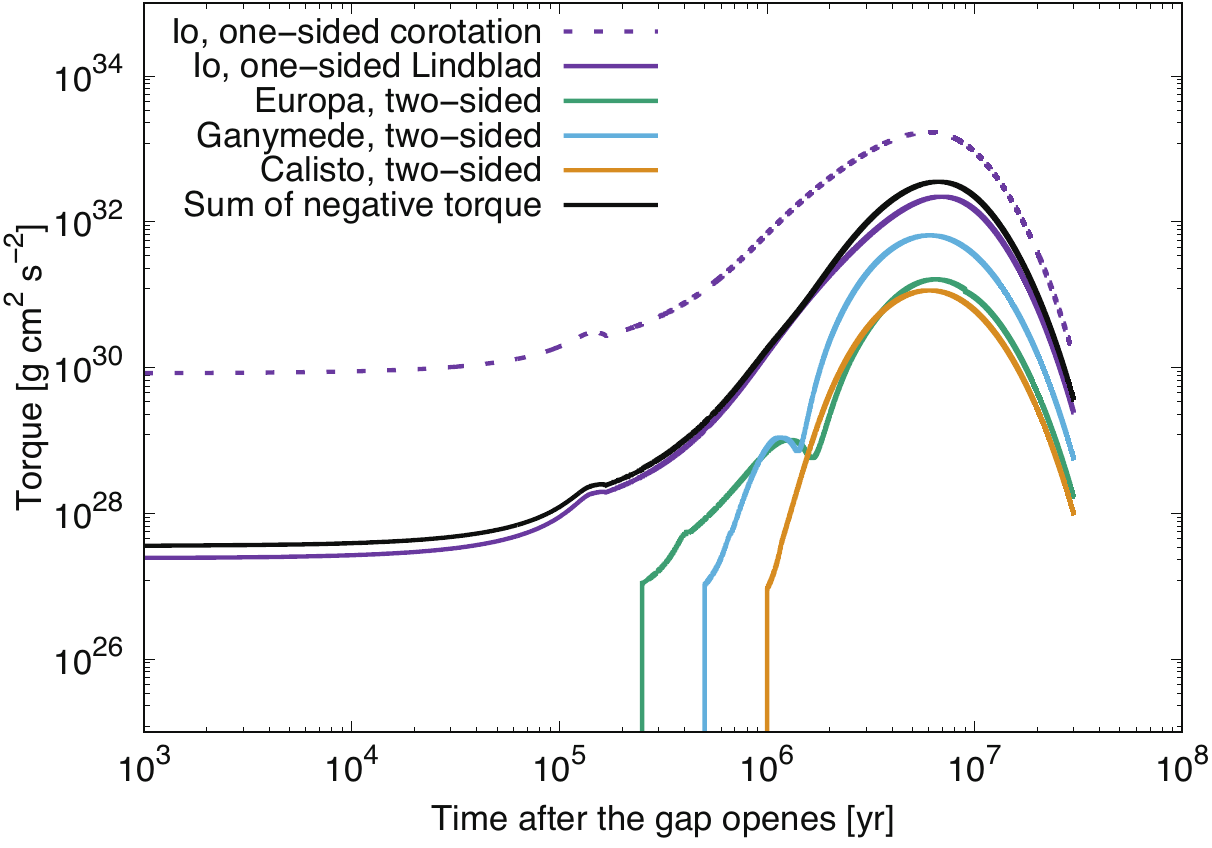}
\caption{{\bf Torque which the satellites receive (Model A).} The solid and dashed purple curves are the Lindblad and corotation one-sided torque that Io (Seed A1) receives, respectively. The green, light blue, and orange solid curves are the two-sided torque of Europa (Seed A2), Ganymede (Seed A3), and Callisto (Seed A4), respectively. The black curve is the sum of the negative torque; the Lindblad one-sided torque of Io and the two-sided torque of the other three satellites. \label{fig:migration}}
\end{figure}

First, we check if Io can stop its inward migration at the edge and move outward together or not. The two-sided (i.e., normal Type I migration) torque that a satellites receives is,
\begin{equation}
\Gamma_{\rm 2s}=\dfrac{1}{2}M_{\rm s}v_{\rm K}v_{\rm t1}.
\label{gamma2s}
\end{equation}
On the other hand, a satellite at the disk edge receives a strong one-sided positive corotation torque that pushes the satellite outward \citep{liu17,rom19}. The one-sided corotation and Lindblad torque is,
\begin{equation}
\Gamma_{\rm 1s,co}=2.46\left(\dfrac{\Sigma_{\rm g}r^{2}}{M_{\rm cp}}\right)\left(\dfrac{\mu_{\rm s}}{h_{\rm g}^{3}}\right)^{1/2}M_{\rm s}(r\Omega_{\rm K})^{2},
\label{gamma1sco}
\end{equation}
and
\begin{equation}
\Gamma_{\rm 1s,Lin}=-0.65\left(\dfrac{\Sigma_{\rm g}r^{2}}{M_{\rm cp}}\right)\dfrac{\mu_{\rm s}}{h_{\rm g}^{3}}M_{\rm s}(r\Omega_{\rm K})^{2},
\label{gamma1sLin}
\end{equation}
respectively \citep{liu17}. Figure \ref{fig:migration} shows that the one-sided corotation torque Io (Seed A1) receives is much larger than the negative one-sided Lindblad torque and the negative two-sided torque of that the other satellites (Seeds B2, B3, and B4) receive, respectively. Since the sum of the negative torque is smaller than the positive torque, even if the other satellites push Io through the chain of the resonance, Io keeps its position on the edge of the cavity. It is also argued that if the migration timescale of the edge (i.e., $3.5~t_{\rm dep}$) is shorter than that of the migration of the satellite by the one-sided corotation torque, the disk edge leaves the satellite there \citep{liu17}. We found that, however, the timescale of Io's migration by the one-sided corotation torque is $\sim10^{3}~{\rm year}$ and this is much shorter than the migration timescale of the edge, $\sim10^{6}~{\rm year}$. Therefore, Io may have moved much further than its current orbit with the disk inner edge moves outward.

However, this outward migration of the inner edge should stop at the corotation radius  $r_{\rm co}$ where the Keplerian frequency of the disk equals to the spin frequency of Jupiter. When $r_{\rm cav}>r_{\rm co}$, there will be two possibilities; the angular momentum will be transferred from Jupiter to the disk and then the gas accretion will stop, or otherwise the corotation radius and the disk edge will move outward together and then the accretion will continue \citep{tak96,liu17}. Although the current corotation radius is $r_{\rm co}\approx2.25~R_{\rm J}$, Jupiter at its time of formation was much larger than it is today \citep{bur97,for11}, and this means that the corotation radius was also larger than the current one if the conservation of the angular momentum of Jupiter is assumed. Considering the transport of the angular momentum from Jupiter to the disk, the angular momentum should have been conserved since the disk disappeared. According to a formation model of Jupiter, the radius of the planet was $\approx1.75~R_{\rm J}$ after its rapid gas accretion and it decreased little by little \citep{lis09}. When the radius of Jupiter is $1.75~R_{\rm J}$, the corotation radius should be $r_{\rm co}\approx4.7~R_{\rm J}$. We can then consider two scenarios of Io formation. In the first one, Io formed around $r\approx4.7~R_{\rm J}$, slightly interior to the $r=5.89~R_{\rm J}$ of our fiducial model, and then moved outward after the disk dissipated. The satellites, especially the inner ones, could move outward by the tidal force from Jupiter \citep{yod81}. The outer ones would be pushed by the inner ones and move outward with them because of the resonance. In this case, the position of the snowline should have been more inside than the fiducial case in this work but this thermal condition could easily be reproduced by another parameter set. The second possibility is that Io was not the innermost satellite. If a body was present at $r=3.7~R_{\rm J}$, Io would have been situated at $r=5.7~R_{\rm J}$ if they were trapped in a 2:1 resonance. This orbit is consistent with that the corotation radius when the radius of Jupiter is $\approx1.5~R_{\rm J}$ and this radius can be achieved during the contraction of Jupiter. The innermost body may have been broken by the tidal force of Jupiter when it has entered inside the Roche limit. Current Io, trapped in the Laplace resonance, actually moves inward little by little because of the tidal dissipation and the innermost body may have also experienced such inward migration \citep{lai09}.

The ionization degree of the disk inner region, $\chi_{\rm e}\equiv n_{\rm e}/n_{\rm n}$, where $n_{\rm e}$ and $n_{\rm n}$ are respectively number density of electron and neutral gas, is also important. Without enough ionization, the disk gas can not couple with the rotating magnetic field of Jupiter and the inner cavity does not open. The angular momentum transfer can occur if the magnetic Reynolds number $R_{\rm m}\equiv VH_{\rm g}/\lambda>1$, where $V$ is the relative velocity between the rotating magnetic field and the disk gas, and $\lambda$ is the magnetic diffusivity \citep{tak96}. We note that this magnetic Reynolds number is $\sim10^{3}$ times larger than the Elsasser number, the critical number for the MRI activation, $\Lambda=v_{\rm Az}^{2}/(\lambda\Omega_{\rm K})$, where $v_{\rm Az}$ is the $z$ component of the Alfv\'{e}n velocity. If Jupiter and its magnetic field spin rigidly at the current speed, $V\approx3\times10^{4}~{\rm m~s^{-1}}$, $H_{\rm g}\approx1.1\times10^{7}~{\rm m}$, and $\lambda\approx0.74/\chi_{\rm e}$ at $r=1~R_{\rm J}$ with the disk temperature is $T=1000~{\rm K}$ \citep{bla94}. The condition for angular momentum transport is then $\chi_{\rm e}\gtrsim10^{-12}-10^{-11}$. \citet{tak96} estimated the ionization degree of the circum-Jovian disk including the effects of galactic cosmic rays and radioactive isotope decay and so on. They assumed the minimum mass disk model by \citet{lun82}, whose gas surface density is $100$ times larger than that of our model, and argued that the ionization degree is $\chi_{\rm e}\lesssim10^{-16}$ at the midplane, $\chi_{\rm e}\sim10^{-15}$ at the altitude of one scale height, and $\chi_{\rm e}\sim10^{-13}-10^{12}$ at three scale heights. If the ionization degree is inversely proportional to the gas surface density, although it is a very rough estimate, the ionization degree in our disk model should be $\chi_{\rm e}\lesssim10^{-14}$ at the midplane, $\chi_{\rm e}\sim10^{-13}$ at the altitude of one scale height, and $\chi_{\rm e}\sim10^{-11}-10^{-10}$ at three scale heights. Therefore, the condition for angular momentum transport can be achieved at the altitude higher than the scale height. A possible mechanism to curve the inner cavity of the gas disk is that the transport of angular momentum (i.e., the radial gas flow) is dominated in the upper region and the quick vertical relaxation to hydrostatic conditions provides an accompanying vertical upward drift of gas (see \citet{tak96}). However, this discussion is based on a very rough estimate, more detailed investigation (e.g., MHD simulations) of the disk inner region should be carried out in future.

Moreover, a recent work by \citet{ara19} found that photophoresis in the CJD stops the inward drift of dust particles near the orbit of Io and carves an inner cavity in a broad range of initial conditions. This result also supports our assumption of the inner cavity strongly.

\subsection{Strength of Turbulence} \label{turbulence}
We have also assumed that the strength of turbulence of the viscous accretion disk is $\alpha=10^{-4}$, consistent with the inability of the MRI to operate in the CJD \citep{fuj14}. From the assumptions of the strength of turbulence and the gas accretion rate, we calculate the gas surface density of the disk. The surface density affects many properties of the system, for example, the disk temperature, the pebble accretion rates, and the migration speeds.

\subsection{Disk Temperature} \label{temperature}
\begin{figure}
\epsscale{1.15}
\plotone{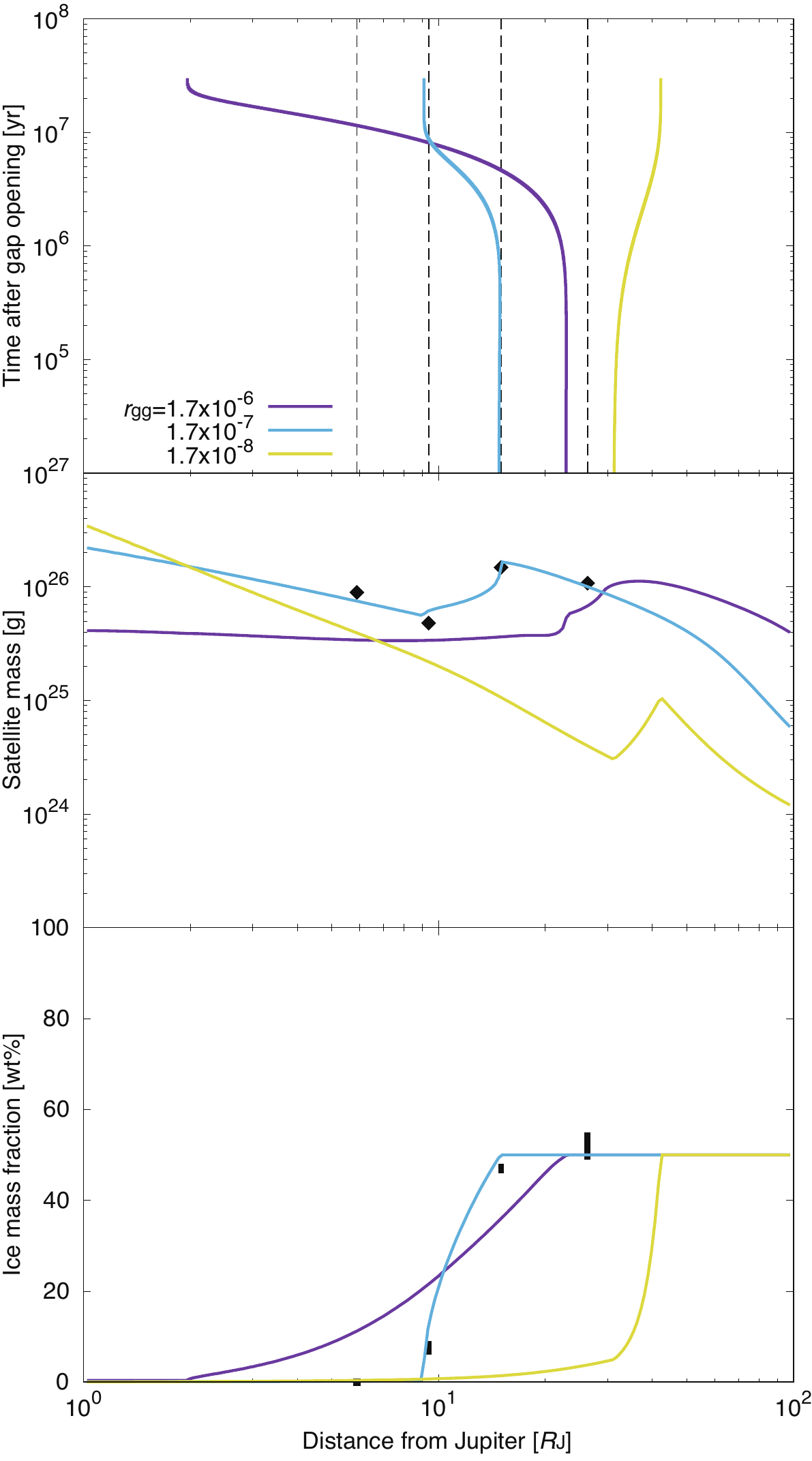}
\caption{{\bf Effects of the thermal condition of the disk.} The top, middle, and bottom panels represent the evolution of the snowline, the final satellite mass, and the final ice mass fractions of satellites, respectively. The color variations represent the difference in the grain-to gas surface density ratio. The black lines and diamonds are the current properties of the Galilean satellites \citep{kus05}. We put seeds of satellites on $1$ to $100~R_{\rm J}$ with the initial mass of $M_{\rm s,start}=3\times10^{23}~{\rm g}$ and fix their positions in this calculation. The dust-to-gas accretion ratio is $x=0.0021$. \label{fig:sdg}}
\end{figure}

To get the disk thermal condition suitable for reproducing the ice mass fractions of the satellites, we needed that Rosseland mean optical depth $\tau$ is around unity at Ganymede's orbit. Figure \ref{fig:sdg} represents the $r_{\rm gg}$ dependences of the snowline, satellite mass, and ice fraction of satellites in the case of the fixed orbits. The top panel shows that when $r_{\rm gg}=1.7\times10^{-7}$, the snowline migrates inward from just inside Ganymede's orbit and stops just inside Europa's. This is because the first term is dominant in the opacity factor $g$ in the first half of its evolution and then the second term becomes dominant as the gas surface density decreases (see Eq. (\ref{g-b})). If $r_{\rm gg}$ is larger than this, the first term is dominant until the almost end, and vice versa. The middle panel shows that the mass distribution of the Galilean satellites can be reproduced only when $r_{\rm gg}=1.7\times10^{-7}$. However, if the gas accretion rate $\dot{M}_{\rm g}$ is smaller than this assumption and the dust-to-gas accretion rate ratio $x$ is correspondingly larger, the snowline will again be near Ganymede and the mass distribution of the satellites will be reproduced. The bottom panel shows that the $r_{\rm gg}$ dependence of the ice mass fraction is large. When $r_{\rm gg}$ is larger than the most suitable case, the slope inside the first position of the snowline is gentler and the ice mass fraction of Europa is larger. On the other hand, when $r_{\rm gg}$ is smaller, the slope is steeper and Europa can not get enough ice. These results can be understood by the top panel.

However, this constraint of marginal optical thickness could be plausible. First, the opacity of the drifting pebbles is consistent with our assumption. In the geometric limit (pebbles larger than the wavelength), the opacity of pebbles is given by $\kappa_{\rm p}\sim1/(\rho_{\rm int,p}a_{\rm p})$, where $\rho_{\rm int,p}$ and $a_{\rm p}$ are the pebble internal density and the pebble radius, respectively. If the pebbles are fluffy, the opacity of pebbles could be $\kappa_{\rm p}\sim1/(10^{-3}\times10)=10^{2}~{\rm cm^{2}~g^{-1}}$ \citep{kat14,shi17}. In this case, the mean optical depth of pebbles is $\tau_{\rm p}=\kappa_{\rm p}\Sigma_{\rm p}\sim10^{2}\times10^{-2}=1$ (see Figure \ref{fig:disk}). This is consistent with the Rosseland mean optical depth in our models, $\tau=450r_{\rm gg}\Sigma_{\rm g}\sim500\times(2\times10^{-7})\times10^{4}\sim1$ (see Eq. (\ref{kappa})). Note that the gas (molecular) opacity ($\sim10^{-5}-10^{-4}~{\rm cm^{2}~g^{-1}}$, consistent with an optical depth of $\sim0.1-1$ with $\Sigma_{\rm g}\sim10^{4}~{\rm g^{1}~cm^{-2}}$) may also affect the disk temperature but the effect should be limited \citep{miz80}. Second, in reality Eqs.(\ref{T4-b}) and (\ref{g-b}) may not be applicable because the dissipation of accretion energy occurs in the disk upper regions if the disk is laminar and wind-driven accreting. This implies that optically thick disks can still have cold midplanes \citep{hir09,mor19}. Under these conditions, our model will also work with higher $r_{\rm gg}$.

Note that, we simply assumed the snowline as the position where the disk temperature is $160~{\rm K}$ after previous models of PPDs, the gas pressure of the CJD is much higher than that of PPDs, the partial water vapor pressure and the condensation temperature should also be higher. In this case, the evolution track of snowline should be shifted closer to Jupiter causing the increase of the ice mass fraction of Europa. Our model will be able to reproduce the ice-depleted Europa again if we chose a larger $\dot{M}_{\rm g}$ but it means a shorter formation timescale and may make the difference between the internal structures of Ganymede and Callisto smaller.

\subsection{A Little Supply of Solid Material to the CJD} \label{material}
One of the strong points compared with previous scenarios is that our slow-pebble-accretion scenario only needs a small amount of solid material (i.e., dust particles and planetesimals). The total amount of dust needed to drive the growth in the slow-pebble-accretion scenario is also modest and is smaller than that for the classical satellitesimal-accretion scenarios \citep{can06,shi17}. Our scenario can reproduce the mass of the current Galilean satellites even with the dust-to-gas accretion ratio is as low as $x=0.0026$, which is smaller than the solar composition by a factor of four. This is consistent with the fact that the gas flowing into the CJD is depleted in solids \citep{can02,tan12}. Because of the gap formation in the CJD caused by Jupiter, only small particles ($\lesssim0.1~{\rm mm}$) can overcome the ensuing gas pressure gradient to end up in the CJD \citep{zhu12}, In addition, the accreted gas should be supplied from high altitude where gravitational settling of larger particles limits the amounts of dust \citep{tan12}. On the other hand, small amounts of particles ($\sim0.1~{\rm mm}$) can form as fragments of the collisions of the solids trapped at the gas pressure maximum \citep{kob12}.

Furthermore, only four planetesimals need to be captured by the CJD. This assumption is consistent with the following facts: after the gap opens in the CSD, such captures only appear for the planetesimals which have high eccentricities in the CSD \citep{fuj13}; planetesimals can form at the gas pressure maximum of the gap and those that are scattered by a large body will have high eccentricities \citep{kob12,ron18}.

\subsection{Properties of the Captured Planetesimals} \label{seeds}
We consider the cases that the planetesimals which have formed in the CJD are captured by the CJD and become the seeds of the satellites. The assumptions of a starting location of $r_{\rm s,start}=50~R_{\rm J}$ with zero eccentricity are plausible. According to a previous work, there are orbits that planetesimals are captured by multiple approaches to Jupiter and become circularized around $\approx50~R_{\rm J}$ (long-lived prograde captured orbits) \citep{sue16}.

We also fix the mass of the captured planetesimals as $3\times10^{23}{\rm g}$ (about $300~{\rm km}$), which is consistent for the properties of the captured orbits \citep{sue16}. These planetesimals are large enough to start growing by efficient pebble accretion (i.e., in the settling regime) for the adopted disk properties (see Eq. (\ref{Mstar})). Smaller planetesimals, on the other hand, will stay small and may be scattered or accreted by large ones. In the settling regime, any difference in the initial mass does not change the final mass because the growth timescale increases with mass (see the lower panel of Figure \ref{fig:peff}). In addition, planetesimals smaller than the assumed value, are likely to end up in higher order resonances (i.e., 3:2 instead 2:1), by virtue of their faster migration (see Eqs. (\ref{vad}), (\ref{Tcrit}), and Table \ref{tab:CMMR}). Therefore, the initial seed mass must be similar or larger than our standard value.

\subsection{Capture-time of the Planetesimals} \label{Callisto}
Each capture time of the planetesimals $t_{\rm cap}$ are assumed to make the dichotomy of the internal structures of the Galilean satellites; Callisto is minimally or only modestly differentiated but the other satellites are fully differentiated. It is difficult to make the dichotomy between the internal structures of Ganymede and Callisto by the difference in the heat of accretion or $^{26}$Al decay in the previous Canup-Ward formation scenario \citep{can06,bar08}. In our slow-pebble-accretion scenario, we found that only Ganymede was differentiated by $^{26}$Al heat provided the seed of Callisto was captured by the disk at a sufficiently late time.

\begin{figure}
\epsscale{1.15}
\plotone{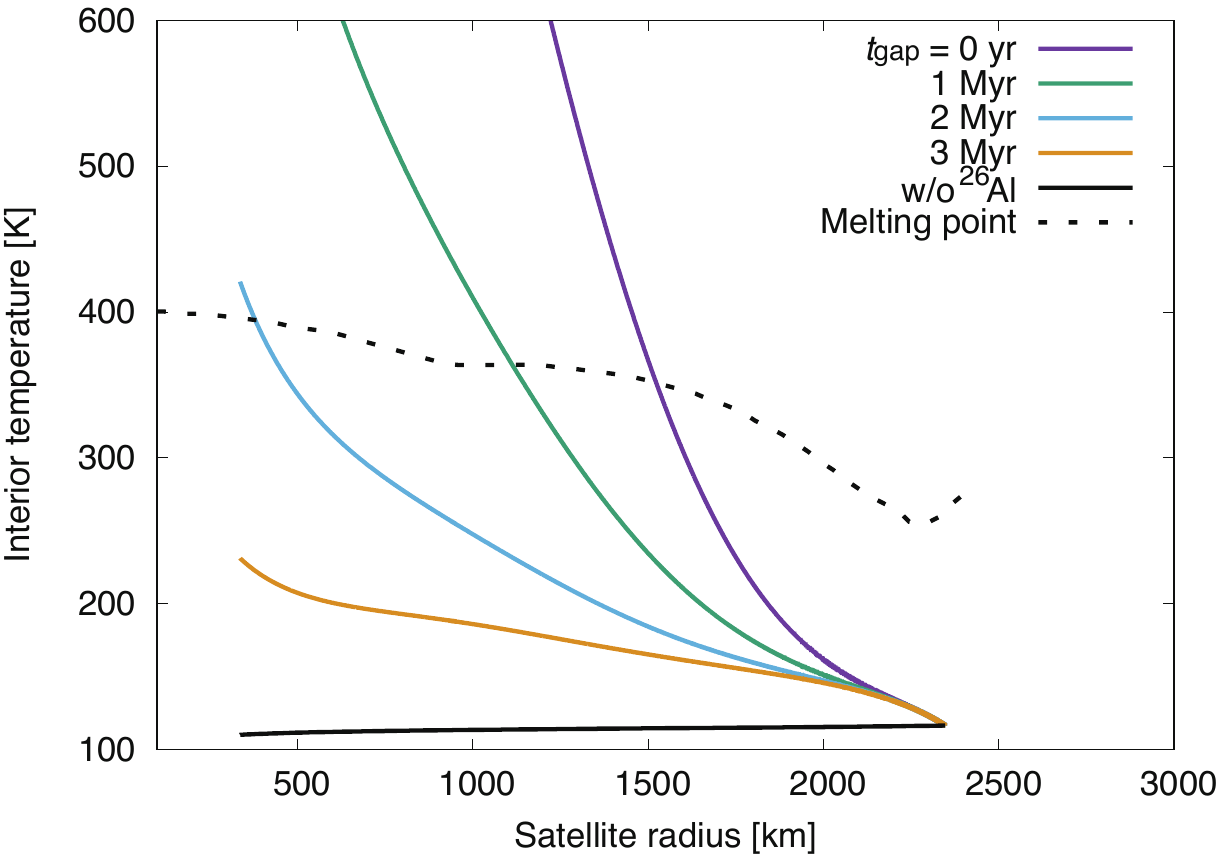}
\caption{{\bf Internal temperature of satellites.} The color variations represent the difference in the gap opening time, in other words, the time that the pebble accretion starts. The black solid curve represents the internal temperature without $^{26}$Al decay heating (only accretion heating). The dotted black curve is the melting point of Callisto \citep{bar08}. We put seeds of satellites on $26.3~R_{\rm J}$ (the current orbit of Callisto) at $t=0,1,2,3~{\rm Myr}$ with the initial mass of $M_{\rm s,start}=3\times10^{23}~{\rm g}$ and fix their positions in this calculation. The dust-to-gas accretion ratio is $x=0.0021$. \label{fig:Callisto}}
\end{figure}

Figure \ref{fig:Callisto} represents the internal temperature of Callisto where its seed is captured at different timing. We fix the position of Callisto at the current orbit in this calculation. Without the $^{26}$Al heat, the surface of Callisto would remain at the same temperature as the disk. Heating by pebble accretion can be neglected, despite the fact the ice-rich (big) pebbles impact the surface at the escape velocity, i.e., similar to planetesimal heating (see Section \ref{sattemp}). However, whereas in the classical planetesimal-driven accretion scenario accretion proceeds quickly, accretion timescales in our pebble-driven scenario are quite long, $10~{\rm Myr}$ (except for Seed B4). This timescale is much longer than the melting critical accretion timescale, $0.6~{\rm Myr}$ \citep{bar08}.

Only in the cases where $^{26}$Al radiogenic heating is considered, the internal temperature exceeds the melting point of water ice. Figure \ref{fig:Callisto} shows that this condition is achieved if most of the $^{26}$Al in the material of Callisto has decayed before its formation starts, in other words, $t_{\rm gap}(=t_{\rm cap})>2~{\rm Myr}$. Also, if Ganymede starts to form by $1.5~{\rm Myr}$ after the formation of CAIs and Callisto starts later than $2.0~{\rm Myr}$, the dichotomy between the two satellites should be explained. Actually, in Model A, we assumed that the capture time of the seeds of Ganymede and Callisto were $1.5$ and $2.0~{\rm Myr}$ after the CAI formation, respectively. In Model B, Callisto's seed formed at the gas pressure bump made by Ganymede at $10.5~{\rm Myr}$, when Ganymede reaches its PIM.

We also note that if there is less $^{26}$Al in pebbles \citep{lar16}, the internal temperature of satellites must be lower. The curve of the internal temperature in $R_{\rm s}$-$T_{\rm s}$ space should approach the black curve (w/o $^{26}$Al) in Figure \ref{fig:Callisto}. In that case, not only Callisto but also Ganymede should be undifferentiated during their formation.

\subsection{Fragmentation of Pebbles} \label{fragmentation}
If the collision speed of pebbles is too fast, they cannot merge and grow larger but fragment. The collision speeds depend on the Stokes number of pebbles so there is a limit of the size of pebbles. We have assumed that the critical fragmentation speeds of rocky and icy pebbles are $v_{\rm cr}=5$ and $50~{\rm m~s^{-1}}$, respectively. As a result, the size (i.e., Stokes number) of pebbles is determined by fragmentation inside the snowline although drift determines it outside the snowline (see Section \ref{pebblegrowth}). In this work, we have adopted the critical fragmentation speeds from numerical studies of collisions of pebbles with $0.1~{\rm \mu m}$ sized monomers \citep{wad09,wad13}. This monomer size is consistent with the typical size of the grains constituting interplanetary dust particles of presumably cometary origin \citep{rie93}. Moreover, the results of previous experimental work are consistent with those of the numerical simulations \citep{blu00,pop00,gun14}. Therefore, our assumption of the critical fragmentation speed is plausible.

\begin{figure}[h]
\epsscale{1.15}
\plotone{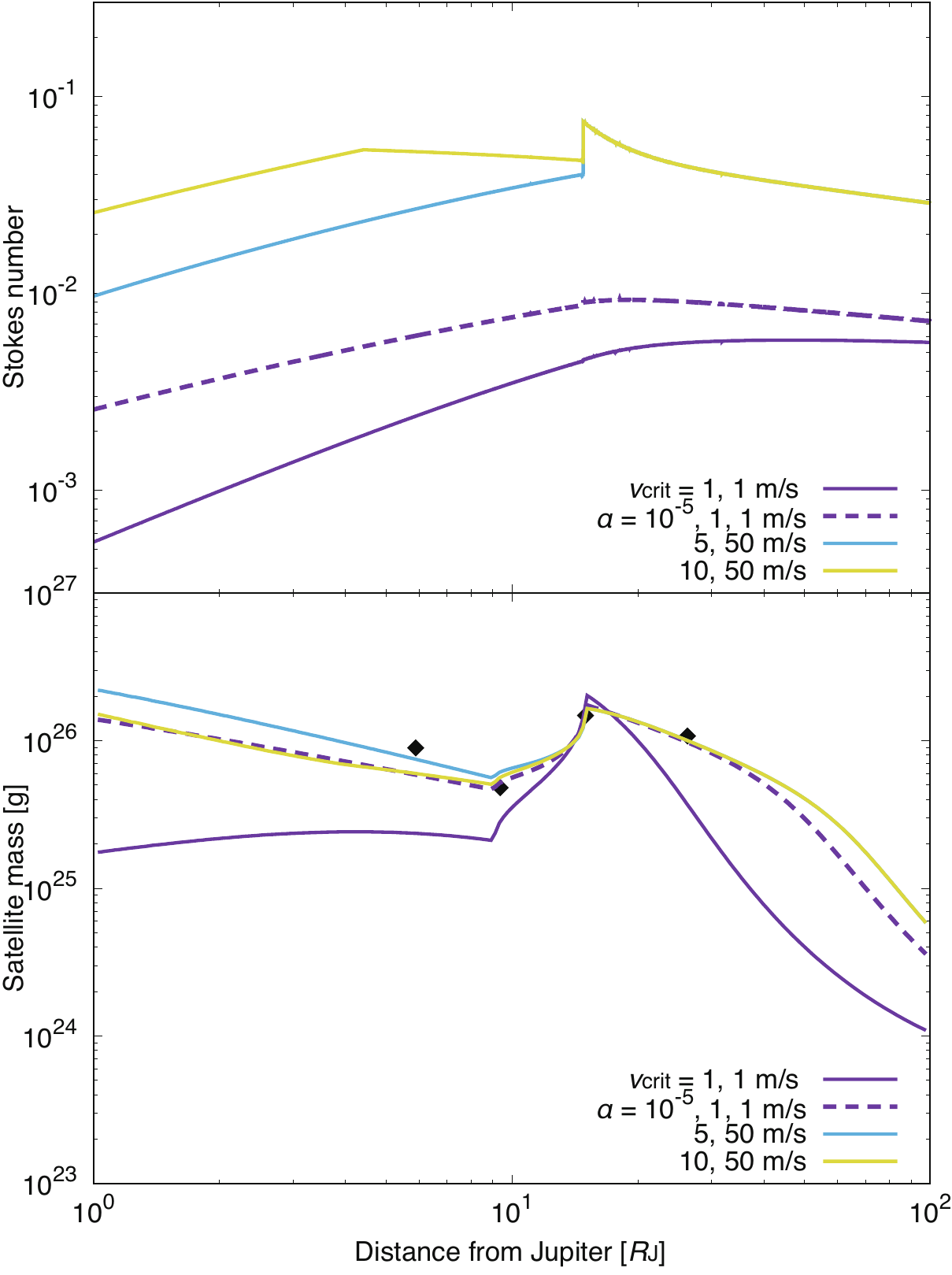}
\caption{{\bf Effects of the fragmentation of pebbles.} The upper and lower panels represent the Stokes number of pebbles ($t=t_{\rm gap}$) and the final mass of satellites, respectively. The purple, light blue, and yelow curves are those with $v_{\rm cr}=1~{\rm m~s^{-1}}$ (rocky and icy pebbles), $v_{\rm cr}=5~{\rm m~s^{-1}}$ (rocky) and $50~{\rm m~s^{-1}}$ (icy), and $v_{\rm cr}=10~{\rm m~s^{-1}}$ (rocky) and $50~{\rm m~s^{-1}}$ (icy), respectively. We put seeds of satellites on $1$ to $100~R_{\rm J}$ with the initial mass of $M_{\rm s,start}=3\times10^{23}~{\rm g}$ and fix their positions in this calculation. The dust-to-gas accretion ratio is $x=0.0021$. The dashed purple curves are the results with $v_{\rm cr}=1~{\rm m~s^{-1}}$ (rocky and icy), $\alpha=10^{-5}$, $r_{\rm gg}=1.7\times10^{-8}$, and $x=0.0011$. \label{fig:vcrit}}
\end{figure}

We note that there is still uncertainty about the fragmentation threshold. A recent experimental work argued that the critical speeds are slower than our assumptions, $\sim1~{\rm m~s^{-1}}$ for both the icy and rocky pebbles \citep{mus19}. On the other hand, \citet{kim15} and \citet{ste19} claimed that the critical fragmentation speed is faster than our fiducial value, $\sim10~{\rm m~s^{-1}}$ for silicate particles. The upper panel of Figure \ref{fig:vcrit} shows that the Stokes number of pebbles with the slower critical speeds (purple) is smaller than the fiducial case (light blue). The size of pebbles is determined by fragmentation not only inside the snowline but also outside. On the other hand, if $v_{\rm cr}=10~{\rm m~s^{-1}}$ for rocky pebbles (yellow), the Stokes number is larger than the fiducial case and is determined by the drift except for the region inside several Jupiter radii. However, the lower panel shows that although the final mass of the satellites with the slow critical speeds (solid purple) is smaller than the fiducial case (light blue), it can be compensated by changing the other parameters. The distribution of the final mass with $v_{\rm cr}=1~{\rm m~s^{-1}}$ (dashed purple) is similar to the fiducial case when $\alpha=10^{-5}$, $r_{\rm gg}=1.7\times10^{-8}$, and $x=0.0011$. The final mass with the higher critical speeds (yellow) is also similar to the fiducial case. Therefore, the effect of the difference of the critical fragmentation speed on our formation scenario is limited.

\subsection{Pebble Isolation Mass} \label{PIM}
In Model B, we define that the PIM of Ganymede is its actual current mass, $1.48\times10^{26}~{\rm g}$. Here, we show this assumption is plausible. An expression of the PIM estimated by 2-D simulations \citet{ata18} is,
\begin{equation}
M_{\rm iso}=h_{\rm g}^{3}\sqrt{37.3\alpha+0.01}\times \left\{1+0.2\left(\dfrac{\sqrt{\alpha}}{h_{\rm g}}\sqrt{\dfrac{1}{{\rm St_{p}}^{2}}+4}\right)^{0.7}\right\}M_{\rm cp}.
\label{Miso}
\end{equation}
We find that this mass is $7.65\times10^{25}~{\rm g}$ with the parameters of Model B, which is about $1.9$ times smaller than the mass of Ganymede. The PIM derived from 2-D simulations can be about 1.5-2 times smaller than the expression by 3-D simulations and the difference can become larger with weak turbulence at least in protoplanetary disks (PPDs) \citep{ata18,bit18,joh19}. Therefore, the PIM of Ganymede must be around the actual mass of the satellite. We also note that Ganymede is the largest satellite so only that satellite should reach its PIM.

\section{Discussions} \label{discussion}
\subsection{Subsequent Evolution of Ganymede's Internal Structure} \label{RTI}
We conclude that Ganymede's interior differentiates into the rocky core and the icy mantle because of the $^{26}$Al heat in the accreted pebbles. However, strictly speaking, it can melt only the region of $1000~{\rm km}$ from the center of Ganymede (see Figure \ref{fig:temp}). Therefore, we should consider the subsequent evolution of the interior, an overturn of the pristine (undifferentiated) icy-rocky crust and the pure icy mantle by the Rayleigh-Taylor (R-T) instability. Here, we consider a three-layered interior structure: (1) the pristine (undifferentiated) icy-rocky crust ($1000-2600~{\rm km}$ from the center), (2) the pure icy mantle ($600-1000~{\rm km}$), and (3) the rocky core ($0-600~{\rm km}$), as the initial condition. We estimate the depth of the layers by assuming the density of the pure ice, rocky core, and pristine crust is $\rho_{\rm ice}=1.4~{\rm g~cm}^{-3}$, $\rho_{\rm rock}=3.0~{\rm g~cm}^{-3}$ and $\rho_{\rm s}=1.9~{\rm g~cm}^{-3}$, respectively.

First, we check whether the Rayleigh-Taylor instability occurs or not. The condition depends on the viscosity of the upper layer, which is the pristine crust in this situation. if the viscosity is lower than the following critical value, the instability occurs with the timescale $t_{\rm RT}$ \citep{rub14},
\begin{equation}
\eta_{\rm crit}=\left[(n-1)^{1/n}\dfrac{C_{\rm L\Delta\rho}}{2n}\right]\left(\dfrac{Z_{0}}{L}\right)^{(n-1)/n}\Delta\rho g_{\rm s}(R_{\rm s})Lt_{\rm RT},
\label{RTcrit}
\end{equation}
where $n$, $C_{\rm L\Delta\rho}$, and $Z_{0}$ are the index of the stress related to the strain rate, a dimensionless quantity determined by the geometry and rheology, and the initial perturbation amplitude, respectively. We assume $n=1.8$, $C_{\rm L\Delta\rho}=0.76$, and $Z_{0}=1~{\rm km}$ (see \citet{rub14}). The difference of density is $\Delta\rho=\rho_{\rm s}-\rho_{\rm ice}=500~{\rm kg~m}^{-3}$ and $g_{\rm s}(R_{\rm s})=0.54~{\rm m~s}^{-1}$ is the gravitational field of satellite at $R_{\rm s}(=1000~{\rm km})$, distance from the center. The lengthscale over which the viscosity varies significantly is
\begin{equation}
L=\dfrac{nRT_{0}}{E_{\rm a}}\dfrac{T_{0}}{|dT_{\rm up}/dz|},
\label{L}
\end{equation}
where $R$, $T_{0}$, and $E_{\rm a}$ are the gas constant, the temperature of at the interface between the two layers, and the activation energy, respectively. We assume them as $T_{0}=370~{\rm K}$ and $E_{\rm a}=49~{\rm kJ~mol}^{-1}$ (see Figure \ref{fig:temp} and \citet{rub14}). According to Figure \ref{fig:temp}, the temperature gradient across the upper crust is $T_{0}/|dT_{\rm up}/dz|=0.14~{\rm K~km}^{-1}$, where $z$ is the vertical distance from the interface. We can then calculate the critical viscosity, $\eta_{\rm crit}=7.6\times10^{13}(t_{\rm RT}/{\rm year})~[{\rm Pa~s}]$. The viscosity of the upper layer should strongly depend on the temperature. According to the Arrhenius functions, the viscosity can be estimated by
\begin{equation}
\eta_{\rm up}=\eta_{\rm ref}\exp\left[A\left(\dfrac{T_{\rm ref}}{T_{\rm up}}-1\right)\right],
\label{Arrhenius}
\end{equation}
where $\eta_{\rm ref}$, $T_{\rm up}$, and $T_{\rm ref}=273~{\rm K}$ are the reference viscosity, the temperature of the upper crust, and the reference temperature (melting point of pure ice), respectively. A constant coefficient $A$ is about $20-25$ and the reference viscosity of pure ice I and pure ice V and VI, are $\eta_{\rm ref}\sim10^{14}~[{\rm Pa~s}]$ and $\sim10^{16}-10^{17}~[{\rm Pa~s}]$, respectively \citep{rub14,sho14}. If the volume rate of rock is smaller than $75\%$, the viscosity of rock-mixed ice is almost the same as that of pure ice \citep{dur10}. The minimum temperature of the upper crust is about $150~{\rm K}$ (see Figure \ref{fig:temp}). If we substitute this temperature for Eq. (\ref{Arrhenius}) and equate $\eta_{\rm up}$ with $\eta_{\rm crit}$, the timescale of the R-T instability is $10^{9}-10^{10}~{\rm year}$ for $A=20$ and $10^{11}-10^{12}~{\rm year}$ for $A=25$. However, if we substitute $T_{\rm up}=170~{\rm K}$, only $20~{\rm K}$ larger, the timescale becomes $10-100$ times shorter. Therefore, if the upper crust was heated up $20~{\rm K}$ by long-lived radiogenic heating or the released gravitational energy of halfway R-T instability itself (see Appendix \ref{RTIheat}), R-T instability has been able to occur in the upper crust of Gaymede within the age of the satellite. One interesting observational fact of Ganymede is that there are two regions on the surface of the satellite: very primitive region ($\sim4~{\rm Ga}$) and relatively newer region ($\sim2~{\rm Ga}$). This characteristic is consistent with our estimate of the evolution of Ganymede's interior; the newer surface had been the inner pure icy mantle and was transported to the surface by the R-T instability around $2~{\rm Ga}$. On the other hand, Callisto's interior did not melt and the surface has been kept as a primitive crust.

\subsection{In-situ Formation of the Seeds} \label{in-situ}
Finally, we highlight the possibility that the seeds of the satellites are not supplied from the CSD but form in-situ around the snowline of the CJD by streaming instability \citep{sch17}. \citet{orm17a} and \citet{sch19} have used this idea towards understanding the properties of the TRAPPIST-1 system. In their model planets consecutively form at the snowline, then migrate inwards. Such in-situ formation of the seeds is in particular attractive to Model B as it would no longer rely on capturing planetesimals. The properties of the satellites in this case will not change from those in the planetesimal-capture case because the seeds must migrate quickly and grow at each current orbital position of the satellites in both of the cases. The number of the Galilean satellites is also naturally explained in Model B. If the seeds forming at the snowline migrate inward and are trapped into 2:1 MMRs one by one, the orbits of the third seed must be around the snowline and no more seeds can form there.

However the main problem with this scenario is that in-situ satellitesimal formation requires dust-to-gas ratios $\sim1$ \citep{car15,yan17}, which is not reached by far because of the very fast pebble drift in the CJD \citep{shi17}. The increase in ice surface density by a factor 3-5 at the snowline \citet{sch17}, would not be enough by far.

\section{Conclusions} \label{conclusions}
We developed a new formation scenario for the Galilean satellites using the pebbles drifting toward Jupiter in the circum-Jovian disk. As the seeds of the satellites, we assumed that only several planetesimals had been captured at given timing. Such conditions should be easier to occur than those assumed in previous satellitesimal-accretion scenarios because it is considered that dust particles in a circumplanetary disk drift toward the central planet before they grow to satellitesimals and a gap structure of planetesimals prevents them from being captured by the disk \citep{fuj13,shi17}.

We first calculated the evolution of the CJD, a gas accretion disk, which was determined by the given gas inflow mass flux reducing exponentially, and the fixed strength of turbulent viscosity. We then calculated 1-D (radial distribution) representative-size evolution of the pebbles including their collisional growth, aerodynamic drag, fragmentation and the distinction of icy/rocky pebbles by the position of snowline. Finally, the growth by accretion of the pebbles and 1-D radial orbital evolution of the seeds including aerodynamic drag, Type I migration, and simple checks of resonance capture were calculated. We simultaneously  calculated their internal thermal evolution by pebble accretion and $^{26}$Al decay heating, and the pebble filtering effects by outer satellite seeds.

In contrast to the previous scenarios which only explain parts of the characteristics of the Galilean system and the scenarios are inconsistent with each other, we found that our new scenario can reproduce the following characteristics simultaneously and consistently with possible assumptions. First, it can reproduce the mass distribution of all the Galilean satellites even in the case of a very small amount of material supply to the CJD. Second, Io, Europa, and Ganymede are captured into 2:1 resonances one by one because the inner cavity opens by the strong magnetic field of Jupiter and halts the migration of Io at the edge of the cavity. Third, Europa accretes small quantities of icy particles in the final phase of its formation because the snowline moves inward as the gas accretion rate onto the CJD decreases. Therefore, Europa's rocky core and icy mantle are explained naturally. The ice mass fractions of the three other satellites are also reproduced because the orbits are inside, respectively outside, the snowline. Finally, our model explains why only Callisto stays undifferentiated and why the other satellites melt by radiogenic heating of $^{26}$Al decay. The difference in the capture time of the planetesimals affects the total $^{26}$Al heat they got and their internal structures but not their final mass because of their slow growth.

We also considered a model that Ganymede reaches its pebble isolation mass (PIM) and Callisto forms by the solid material trapped at the gas pressure maximum created by Ganymede. In this case, the evolution of the inner three satellites are almost same with the first model and the mass of Callisto is also reproduced. The orbits of Ganymede and Callisto are so close that Callisto could be scattered outward after the CJD disappears, resulting in Callisto's unique orbit not trapped in any resonance. The unmelted Callisto is also reproduced and the dichotomy between the internal temperatures of the two satellites is larger than that in the first model. This is because the interior of Callisto is only heated up by (weak) accretion heating and $^{26}$Al heat is negligible due to the late start of growth.

\acknowledgments
We thank Yasuhito Sekine for valuable discussion. This work is supported by JSPS KAKENHI JP16J09590, 15H02065 and 16K17661, and MEXT KAKENHI 18H05438. C.W.O is supported by the Netherlands Organization for Scientific Research (NWO; VIDI project 639.042.422). Part of this research was initiated during his stay as a visiting Professor at the Tokyo Institute of Technology. This work has been carried out within the frame of the National Centre for Competence in Research PlanetS supported by the Swiss National Science Foundation (SNSF). Y.S. acknowledges the financial support of the SNSF.

\appendix
\section{Gas Pressure Maximum formed by Ganymede} \label{maximum}
In Model B, we considered that Callisto formed at the gas pressure maximum created by Ganymede. The local gas surface density around Ganymede can be given by \citep{kob12}
\begin{equation}
\Sigma_{\rm g,local}=\Sigma_{\rm g}\exp[-(r-r_{\rm G}/l)^{-3}],
\label{Sigmaglocal}
\end{equation}
where $r_{\rm G}=14.8~R_{\rm J}$ is the orbital radius of Ganymede (in Model B) and
\begin{equation}
l=\left[\dfrac{8}{81\pi}\dfrac{r_{\rm G}^{2}\Omega_{\rm K}(r_{\rm G})}{\nu}\left(\dfrac{M_{\rm G}}{M_{\rm cp}^{2}}\right)^{2}\right]^{1/3}r_{\rm G},
\label{l}
\end{equation}
where $M_{\rm G}=1.48\times10^{26}~{\rm g}$ is the mass of Ganymede (in Model B). Then the local ratio of the pressure gradient and gravity is
\begin{equation}
\eta_{\rm local}=\dfrac{c_{\rm s}^{2}}{2r^{2}\Omega_{\rm K}^{2}}\left[p+\dfrac{q}{2}+\dfrac{3}{2}-\dfrac{3l^{3}r}{(r-r_{\rm G})^{4}}\right],
\label{etalocal}
\end{equation}
where $p$ and $q$ are the $r$ exponents of the gas surface density and temperature (i.e., $\Sigma_{\rm g}\propto r^{-p}$ and $T_{\rm d}\propto r^{-q}$). Note that in \citet{kob12}, $q$ is defined as the $r$ exponents of the sound speed (see the paper for further explanations about the above equations). At the pressure maximum, $\eta_{\rm local}=0$, so we solve this equation and gain the position of the pressure maximum. When $T=150~{\rm K}$ and $M_{\rm cp}=M_{\rm J}$, and the other parameters are consistent with those in Model B, we find that the pressure maximum is at $r=17.0~R_{\rm J}$ and it is almost independent on these parameters. Therefore, in Model B, we fix the orbital radius of Seed B4 as $r=17.0~R_{\rm J}$.

\section{Heat Released by R-T Instability} \label{RTIheat}
One possible heat source to melt the whole interior of the satellite is the potential energy released during the overturn of the crusts by the R-T instability. We roughly estimate the heat released by the instability and how much the internal temperature rises. If the pure ice mantle (consistent with the half mass of the differentiated region, $M_{\rm ice}(=M_{\rm rock})=4\times10^{21}~{\rm kg}$) is lifted up to the surface of the satellite, the thickness of the pure ice crust will be $h_{\rm ice}=120~{\rm km}$. The mass of the pure ice (and rock) is much smaller than the whole mass of Ganymede ($M_{\rm G}$), the increase in the temperature of the undifferentiated pristine crust uniformly heated up by the released potential energy ($\Delta Q_{\rm RT}$) can be roughly estimated by the following equations,
\begin{equation}
\begin{split}
\Delta T_{\rm RT}=&\dfrac{\Delta Q_{\rm RT}}{C_{\rm p}M_{\rm G}} \\
\approx&\dfrac{GM_{\rm ice}}{C_{\rm p}(R_{\rm G}-h_{\rm ice})}\left(\dfrac{\rho_{\rm s}}{\rho_{\rm ice}}-1\right)\left(1-\dfrac{R_{\rm rock}^{2}\rho_{\rm rock}}{(R_{\rm G}-h_{\rm ice})^{2}\rho_{\rm s}}\right) \\
\approx&20~[{\rm K}],
\end{split}
\label{DeltaTRT}
\end{equation} 
where $R_{\rm G}$ is the radius of Ganymede. This temperature increase is, unfortunately, not enough for the whole differentiation of the interior of the satellite. To melt the whole region, about $100~{\rm K}$ increase of the temperature would be needed (see Figure \ref{fig:temp}). However, we neglect the two kinds of heat provided by (1) the released potential energy during the differentiation of the region of $1000~{\rm km}$ from the center and (2) the central captured planetesimal which has been heated up by $^{26}$Al decay \citep{wak11}. Such heat may be transported with the lifting up pure icy mantle and so be able to contribute to differentiate the region remaining unmelted.

\subsection{Parameters, Constants, and Variables} \label{variables}
We summarize the parameters and constants used in this work in Table \ref{tab:parameter}. We also summarize the valuables in Table \ref{tab:variables1} and \ref{tab:variables2}.

\begin{table*}
\centering
\caption{Parameters and constants}
\medskip
\begin{threeparttable}
\begin{tabular}{lll} \hline
 & Value\tnote{1} & Description \\ \hline \hline
$t_{\rm gap}$ & $0, {\bf 1.0}, 2.0, 3.0~{\rm Myr}$ & Gap opening time \\
$\alpha$ & $10^{-5}, {\bf 10^{-4}}$ & Strength of turbulence \\
$x$ & $0.0011, 0.0021, {\bf 0.0026}$ & Dust-to-gas accretion rate ratio \\
$r_{\rm gg}$ & $1.7\times10^{-8}, {\bf 1.7\times10^{-7}}, 1.7\times10^{-6}$ & Grain-to-gas surface density ratio \\
$t_{\rm cap}$ & ${\bf 1.0, 1.25, 1.5, 2.0}~{\rm Myr}$ & Capture time \\
$v_{\rm cr}$ & $1, 1$ / ${\bf 5, 50}$ / $10, 50~{\rm m~s^{-1}}$ & Critical fragmentation speed of rocky or icy pebbles \\
$M_{\rm J}$ & $1.90\times10^{30}~{\rm g}$ & Jupiter mass \\
$\dot{M}_{\rm g,gap}$ & $0.2~M_{\rm J}~{\rm Myr}^{-1}$ & Initial gas accretion rate \\
$t_{\rm dep}$ & $3\times10^{6}~{\rm yr}$ & Gas depletion timescale \\
$R_{\rm J}$ & $7.15\times10^{9}~{\rm cm}$ & Jupiter radius \\
$r_{\rm cav}$ & $5.89~R_{\rm J}$ & Width of the magnetospheric cavity (see Eq. (\ref{rcav})) \\
$M_{\rm G}$ & $1.48\times10^{26}~{\rm g}$ & Mass of Ganymede \\
$m_{\rm r}$ & $0.5$ & Rock mass fraction outside the snowline \\
$r_{\rm s,start}$ & $50~R_{\rm J}$ & Initial (captured) position of satellites (seeds) \\
$M_{\rm s,start}$ & $3\times10^{23}~{\rm g}$ & Initial mass of satellites (seeds) \\
$t_{\rm fin}$ & $3\times10^{7}~{\rm yr}$ & Time of the end of the formation \\
$m_{\rm g}$ & $3.9\times10^{-24}~{\rm g}$ & Mean molecular mass of gas \\
$C_{\rm D}$ & $0.5$ & Drag coefficient \\
$C_{\rm MMR}$ & Table \ref{tab:CMMR} & Capture coefficient \\
$\lambda_{26}$ & $9.68\times10^{-7}~{\rm yr^{-1}}$ & Decay rate of $^{26}$Al \\
$q_{26,0}$ & $1.82\times10^{-7}~{\rm W~kg^{-1}}$ & Initial heating rate by $^{26}$Al \\
$C_{\rm p}$ & $1400~{\rm J~kg^{-1}~K^{-1}}$ & Specific heat (for Ganymede and Callisto) \\
$\rho_{\rm s}$ & $1.9~{\rm g~cm^{-3}}$ & Satellite density (for Ganymede and Callisto) \\
$\rho_{\rm int,p}$ & $\sim10^{-3}~{\rm g~cm^{-3}}$ & Internal density of fluffy pebbles \\
$a_{\rm p}$ & $\sim10~{\rm cm}$ & Radii of fluffy pebbles \\
$\rho_{\rm ice}$ & $1.4~{\rm g~cm^{-3}}$ & Density of pure ice \\
$\rho_{\rm rock}$ & $3.0~{\rm g~cm^{-3}}$ & Density of the rocky core \\
$n$ & $1.8$ & Index of the stress related to the strain rate \\
$C_{\rm L\Delta\rho}$ & $0.76$ & Dimensionless quantity determined by the geometry and rheology \\
$Z_{0}$ & $1~{\rm km}$ & Initial perturbation amplitude \\
$T_{0}$ & $370~{\rm K}$ & Temperature at the interface between the two layers \\
$E_{\rm a}$ & $49~{\rm kJ~mol}^{-1}$ & Activation energy \\
$T_{\rm ref}$ & $273~{\rm K}$ & Reference temperature \\
$A$ & $20, 25$ & Coefficient in the Arrhenius functions \\
$B_{\rm cp}$ & $\approx4, {\bf 40}~{\rm Gauss}$ & Strength of the magnetic field of the central planet \\
$t_{\rm iso,G}$ & $9.54~{\rm Myr}$\tnote{2} & Time that Seed B3 reaches its PIM \\ \hline
\end{tabular}
\begin{tablenotes}
\item[1] The boldface shows the fiducial case.
\item[2] A result of Model B
\end{tablenotes}
\end{threeparttable}
\label{tab:parameter}
\end{table*}

\begin{table}
\centering
\caption{Variables} 
\medskip
\begin{tabular}{lll} \hline
 & Equation & Description \\ \hline \hline
$t$ & - & Time after the formation of CAIs \\
$r$ & - & Distance from Jupiter \\
$M_{\rm in}$ & - & Mass of the inner satellite \\
$M_{\rm out}$ & - & Mass of the outer satellite \\
$T_{\rm in}$ & - & Orbital period of the inner satellite \\
$n_{\rm e}$ & - & Number density of electron \\
$n_{\rm n}$ & - & Number density of neutral gas \\
$V$ & - & Relative velocity between the rotating magnetic field and the disk gas \\
$\lambda$ & - & Magnetic diffusivity \\
$v_{\rm Az}$ & - & $z$ component of the Alfv\'{e}n velocity \\
$M_{\rm ice}$ & - & Mass of the pure ice mantle of Ganymede \\
$M_{\rm rock}$ & - & Mass of the rocky core of Ganymede \\
$M_{\rm rock}$ & - & Thickness of the pure ice crust of Ganymede \\
$\Delta Q_{\rm RT}$ & - & Released potential energy by R-T instability \\
$R_{\rm G}$ & - & Radius of Ganymede \\
$\Omega_{\rm K}$ & $\sqrt{GM_{\rm cp}/r^{3}}$ & Kepler angular velocity ($G$: Gravitational constant) \\
$c_{\rm s}$ & $\sqrt{k_{\rm B}T_{\rm d}/m_{\rm g}}$ & Sound speed ($k_{\rm B}$: Boltzmann constant) \\
$\tau$ & $\kappa\Sigma_{\rm g}$ & Rosseland mean optical depth \\
$v_{\rm K}$ & $r\Omega_{\rm K}$ & Kepler velocity \\
$\rho_{\rm g}$ & $\Sigma_{\rm g}/(\sqrt{2\pi}H_{\rm g})$ & Gas density at the midplane \\
$p$ & $\Sigma_{\rm g}\propto r^{-p}$ & $r$ exponent of the gas surface density \\
$q$ & $T_{\rm d}\propto r^{-q}$ & $r$ exponent of the temperature \\
$\mu_{\rm s}$ & $M_{\rm s}/M_{\rm cp}$ & Satellite-to-central planet mass ratio \\
$h_{\rm p}$ & $H_{\rm p}/r$ & Pebble aspect ratio \\
$H_{\rm g}$ & $c_{\rm s}/\Omega_{\rm K}$ & Gas scale hight \\
$t_{\rm grow}$ & $M_{\rm s}/(dM_{\rm s}/dt)$ & Growth timescale \\
$h_{\rm g}$ & $H_{\rm g}/r$ & Gas aspect ratio \\
$R_{\rm col}$ & $2\pi\eta P_{\rm eff}$ & Dimensionless accretion rate \\
$v_{\rm esc}$ & $\sqrt{2GM_{\rm s}/R_{\rm s}}$ & Escape velocity \\
$v_{\rm set}$ & $g_{\rm s}t_{\rm stop}$ & Settling velocity, Terminal velocity \\
$g_{\rm s}$ & $GM_{\rm s}/R_{\rm s}^{2}$ & Gravitational acceleration at the surface of the satellite \\
$t_{\rm stop}$ & ${\rm St_{p}}/\Omega_{\rm K}$ & Stopping time of pebbles \\
$r_{\rm H,G}$ & $(M_{\rm G}/(3M_{\rm J}))^{1/3}r_{\rm G}$ & Hill radius of Ganymede \\
$\chi_{\rm e}$ & $n_{\rm e}/n_{\rm n}$ & Ionization degree \\ 
$R_{\rm m}$ & $VH_{\rm g}/\lambda$ & Magnetic Reynolds number \\ 
$\Lambda$ & $v_{\rm Az}^{2}/(\eta_{\rm m}\Omega_{\rm K})$ & Elsasser number \\
$\kappa_{\rm p}$ & $\sim1/(\rho_{\rm int,p}a_{\rm p})$ & Opacity of pebbles \\
$\tau_{\rm p}$ & $\kappa_{\rm p}\Sigma_{\rm p}$ & Mean optical depth of pebbles \\
$\Delta\rho$ & $\rho_{\rm s}-\rho_{\rm ice}$ & Difference of density \\ \hline
\end{tabular}
\label{tab:variables1}
\end{table}

\begin{table}
\centering
\caption{Variables}
\medskip
\begin{threeparttable}
\begin{tabular}{lll} \hline
 & Equation & Description \\ \hline \hline
$M_{\rm s}$ & Eq. (\ref{Ms}) & Satellite mass \\
$\dot{M}_{\rm g}$ & Eq. (\ref{Mdotg}) & Gas accretion rate \\
$\Sigma_{\rm g}$ & Eq. (\ref{Sigmag}) & Gas surface density \\
$T_{\rm d}$ & Eq. (\ref{T4-b}) & Disk midplane temperature \\
$g$ & Eq. (\ref{g-b}) & Opacity factor \\
$\kappa$ & Eq. (\ref{kappa}) & Rosseland mean opacity \\
$\dot{M}_{\rm p}$ & Eq. (\ref{Mdotp}) & Pebble mass flux \\
${\rm St_{p}}$ & Eq. (\ref{stdrift}) & Stokes number of pebbles (determined by drift) \\
$v_{\rm pp}$ & Eq. (\ref{vpp}) & Pebble-to-pebble relative velocity \\
$v_{\rm r}$ & Eq. (\ref{vrdash}) & Drift velocity of pebbles \\ 
$v_{\rm t}$ & Eq. (\ref{vtdash}) & Relative velocity of pebbles driven by turbulence \\
$\eta$ & Eq. (\ref{eta}) & Ratio of the pressure gradient force to the gravity of Jupiter \\
${\rm St_{p}}$ & Eq. (\ref{stfrag}) & Stokes number of pebbles (determined by fragmentation) \\
$\Sigma_{\rm p}$ & Eq. (\ref{Sigmap}) & Pebble surface density \\
$P_{\rm eff}$ & Eq. (\ref{Peff}) & Pebble accretion efficiency \\
$H_{\rm p}$ & Eq. (\ref{Hp}) & Pebble scale height \\
$\Delta v$ & Eq. (\ref{Deltav}) & Pebble-satellite relative velocity \\
$\dot{M}_{\rm p,in}$ & Eq. (\ref{Mdotpin}) & Pebble mass flux inside a satellite \\
$M_{\ast}$ & Eq. (\ref{Mstar}) & Satellite mass that the effective pebble accretion starts \\
$dM_{\rm s}/dt$ & Eq. (\ref{dMsdt}) & Mass growth rate of Callisto in Model B \\
$T_{\rm s}$ & Eq. (\ref{Ts}) & Temperature of the satellite surface \\
$u_{\rm i}$ & Eq. (\ref{ui}) & Pebble-satellite collision velocity \\
$\Delta T_{\rm fin}$ & Eq. (\ref{DeltaTfin}) & Increase in the satellite internal temperature \\
$T_{\rm fin}$ & Eq. (\ref{Tfin}) & Final satellite internal temperature \\
$v_{\rm ad}$ & Eq. (\ref{vad}) & Aerodynamic drag migration velocity \\
${\rm St_{s}}$ & Eq. (\ref{Sts}) & Stokes number of satellites \\
$v_{\rm mig}$ & Eq. (\ref{vt1}) & Type I migration velocity \\
$t_{\rm crit}$ & Eq. (\ref{Tcrit}) & Critical migration timescale for resonance capture \\
$b_{\rm t1}$ & * & Migration constant \\
$\Gamma_{\rm 2s}$ & Eq. (\ref{gamma2s}) & Two-sided torque \\
$\Gamma_{\rm 1s,co}$ & Eq. (\ref{gamma1sco}) & One-sided corotation torque \\
$\Gamma_{\rm 1s,Lin}$ & Eq. (\ref{gamma1sLin}) & One-sided Lindblad torque \\ 
$M_{\rm iso}$ & Eq. (\ref{Miso}) & Pebble isolation mass \\
$\eta_{\rm crit}$ & Eq. (\ref{RTcrit}) & Critical viscosity to R-T instability \\ 
$L$ & Eq. (\ref{L}) & Lengthscale over which the viscosity varies significantly \\
$\eta_{\rm up}$ & Eq. (\ref{Arrhenius}) & Viscosity of the upper layer \\
$\Sigma_{\rm g,local}$ & Eq. (\ref{Sigmaglocal}) & Local gas surface density around Ganymede \\
$l$ & Eq. (\ref{l}) & - \\ 
$\eta_{\rm local}$ & Eq. (\ref{etalocal}) & Local ratio of the pressure gradient and gravity \\
$\Delta T_{\rm RT}$ & Eq. (\ref{DeltaTRT}) & Increase in temperature by R-T instability \\ \hline
\end{tabular}
\begin{tablenotes}
\item * Eq. (10) of \citet{ogi15}
\end{tablenotes}
\end{threeparttable}
\label{tab:variables2}
\end{table}

\clearpage
\bibliography{shibaike2019}

\end{document}